\documentclass{aa}
\usepackage{aas_macros}
\usepackage{mathtools}
\usepackage[normalem]{ulem}
\DeclareRobustCommand{\vvec}[1]{\boldsymbol{#1}}
\newcommand{\vvers}{\mathbf{\hat{e}}}
\DeclareRobustCommand{\eps}{\varepsilon}
\renewcommand{\vec}{\mathbf}
\DeclareRobustCommand{\hnabla}{\hat{\nabla}}
\DeclareRobustCommand{\Pp}{\mathbb{P}_\perp}
\DeclareRobustCommand{\Pa}{\mathbb{P}_1}
\DeclareRobustCommand{\Pb}{\mathbb{P}_2}
\DeclareRobustCommand{\Pc}{\mathbb{P}_3}
\usepackage{graphicx}
\usepackage{txfonts}
\usepackage{siunitx}
\usepackage[colorlinks=true, linkcolor=blue, citecolor=blue, filecolor=blue, urlcolor=blue]{hyperref}
\begin{document}
\title{The effects of expansion and turbulence on the interplanetary evolution of a magnetic cloud}
\subtitle{}
\author{M. Sangalli\inst{1}, A. Verdini\inst{1}, S. Landi\inst{1} \and E. Papini\inst{2}}
\institute{%
	Dipartimento di Fisica e Astronomia, Università degli Studi di Firenze, Via G. Sansone 1, 50019 Sesto Fiorentino, Italy\\
	\email{mattia.sangalli@unifi.it}
	\and
	Istituto di Astrofisica e Planetologia Spaziali, Istituto Nazionale di Astrofisica, Via del Fosso del Cavaliere 100, 00133 Roma, Italy
}
\authorrunning{M. Sangalli {\em et al.}}
\date{Received 17 March, 2025; Accepted 13 May, 2025}
\abstract{
Coronal mass ejections (CMEs) represent the most extreme solar products, showing complex and dynamic structures when detected in situ.
They are often preceded by a shock and carry a magnetic cloud organised as a flux rope, surrounded and permeated by turbulent fluctuations, and whose radial size expands during propagation.
We investigate the internal dynamics of the 2D section of a cylindrical flux rope propagating at constant velocity in the spherically expanding solar wind, employing the expanding box model, which allows for high spatial resolution. Our setting is simplified, with uniform and non-magnetised solar wind, to which we superpose turbulent fluctuations.
We find that the spherically expanding geometry alone perturbs the flux rope equilibrium, producing a radial head-tail velocity profile and a radial size increase.
The ratio between the expansion and Alfv{\'e}n timescales, associated respectively to propagation and internal crossing time, controls the resistance to transverse stretching and the increase of the flux rope radial extent; the plasma beta controls the overall size of the structure.
Turbulent fluctuations mainly affect the flux rope transverse structure, spreading its axial field at distances comparable to its size; on the contrary, dynamics along the radial direction remains coherent and the increase in radial size is still consistently observed.
We validate our results by comparison with statistical observations and dimensionless estimates, such as the expansion parameter
and the radial size scaling exponent, suggesting that the ratio between internal and propagation timescales might help in better classifying different kinds of radial expansions for flux ropes.
}
\maketitle
%
%
\section{Introduction}
Coronal mass ejections (CMEs) are large-scale eruptive events taking place in the Sun's outer atmosphere, involving the release of plasma and magnetic field \citep{webb2012LRSP....9....3W}. The ejected material propagates in the solar wind, becoming an Interplanetary CME (ICME). ICMEs often display a complex multipart structure, possibly including a leading shock wave \citep[see e.g.][]{gosling1975SoPh...40..439G, sheeley1985JGR....90..163S}, a compressed and turbulent sheath region \citep{kilpua2017LRSP...14....5K} and an enhanced magnetic field (magnetic cloud/ejecta) \citep[see e.g.][]{klein1982JGR....87..613K, burlaga1988JGR....93.7217B}. The latter can be thought of as a magnetic flux rope, with both extremities connected to the Sun \citep[see e.g.][]{burlaga1988JGR....93.7217B, bothmer1998AnGeo..16....1B}.

Coronal mass ejections drive the most intense geomagnetic storms \citep{gosling1991JGR....96.7831G, webb2000JGR...105.7491W, huttunen2004AnGeo..22.1729H, zhang2007JGRA..11210102Z, richardson2012JSWSC...2A..01R} and their sheath regions play a key role in further enhancing the geoeffectivity \citep{huttunen2004AnGeo..22.1729H, kilpua2017LRSP...14....5K}.

Turbulence is found not only in the background solar wind \citep{goldstein1995ARA&A..33..283G,  bruno2013LRSP...10....2B}, but also in ICMEs, both in magnetic clouds \citep[e.g.][]{ruzmaikin1997JGR...10219753R, leamon1998GeoRL..25.2505L, manoharan2000ApJ...530.1061M, sorriso-valvo2021ApJ...919L..30S} and in sheath regions \citep[e.g.][]{kennel1982JGR....87...17K,  kataoka2005GeoRL..3212103K, kilpua2017LRSP...14....5K}.

Stronger turbulence is usually found in sheath regions \citep{good2020ApJ...893..110G, kilpua2020AnGeo..38..999K, sorriso-valvo2021ApJ...919L..30S, marquezrodriguez2023SoPh..298...54M}, but magnetic clouds also show a consistent fraction of energy in the fluctuations \citep{leamon1998GeoRL..25.2505L}, with a wide range of spectral indices \citep{borovsky2019JGRA..124.2406B} and regardless of the subtraction of the large-scale flux rope field \citep{good2023ApJ...956L..30G}.
Recently \citet{pezzi2024A&A...686A.116P} have performed 3D compressible MHD simulations of turbulence interacting with a large-scale magnetic flux rope, showing that the flux rope inhibits the cascade of weak fluctuations, but is overcome by strong enough ones.

Moreover, multi-spacecraft analyses have shown that ICMEs have a fine-scale structure, with magnetic coherence lengths of both magnetic ejecta \citep{lugaz2018ApJ...864L...7L} and sheath regions \citep{ala-lahti2020JGRA..12528002A} being smaller than the ICME size: magnetic scale lengths estimated at $\SI{1}{AU}$ can be as small as $\SI{0.024}{AU}$ (ejecta) and $\SI{0.06}{AU}$ (sheath), whereas ICMEs have estimated sizes around $\SI{0.3}{AU}$.
A number of factors might explain the presence of this meso-scale structure, including their properties upon release in the coronal environment, their internal evolution and the interaction with the surrounding medium during heliospheric propagation.

During their journey in the heliosphere, ICMEs increase their transverse and radial size, as already apparent in remote sensing observations \citep[see e.g.][]{patsourakos2010A&A...522A.100P}.
While the expansion in the transverse direction can be explained as a consequence of the spherical flow of the ICME, its expansion in the radial direction is thought to originate from the magnetic over-pressure with respect to the ambient solar wind
\citep{demoulin2009A&A...498..551D}. The increase in the radial size was first observed indirectly in magnetic clouds \citep[][]{kumar1996JGR...10115667K, bothmer1998AnGeo..16....1B} and more generally in ICMEs \citep[][]{liu2005P&SS...53....3L, wang2005JGRA..11010107W}; it was then found as an almost linear head-tail proton velocity difference \citep{lepping2003SoPh..212..425L, lepping2008AnGeo..26.1919L, jian2008SoPh..250..375J}.
Statistical studies find the radial size $S$ to increase with heliocentric distance $R$ as a power law $S(R) \propto R^{\alpha_R}$, giving the scaling index $\alpha_R$ as a ``global'' estimate of radial expansion, whereas the velocity profile gives a local estimate of the physical dilation the plasma is experiencing.
Assuming the internal Alfv{\'e}n time to be much shorter than the travel time, along with a local cylindrical geometry assumption, \citet{demoulin2008SoPh..250..347D} linked the local velocity profile to the power-law scaling exponent by means of a non-dimensional expansion parameter.
Such locally estimated scaling exponents for unperturbed magnetic clouds at $\SI{1}{AU}$ are around unity ($0.81\pm0.19$ in \citet{demoulin2010AIPC.1216..329D}, $0.91\pm0.23$ in \citet{gulisano2010A&A...509A..39G}).
However, further studies using aligned radial measurements suggest quite weak correlations between global and local expansion estimates \citep{lugaz2020ApJ...899..119L}.

One of the main tools for investigating the evolution of ICMEs is represented by global numerical simulations. Numerical models such as WSA-ENLIL+Cone \citep{odstrcil2004JGRA..109.2116O} and EUHFORIA \citep{pomoell2018JSWSC...8A..35P} can reproduce complex heliospheric configurations including multiple CMEs and structured solar wind (see also \citet{manchester2017SSRv..212.1159M} and references therein).
Despite capturing many important features of the 2.5D and 3D heliospheric and interplanetary evolution of CMEs, such as the aforementioned transverse and radial expansion, global simulations usually lack the numerical resolution to resolve the ICME fine structure and the physical processes occurring at those scales, like turbulence and magnetic reconnection.

Here, we present a novel approach to simulate the interplanetary evolution of a CME with an embedded magnetic cloud. We employ a semi-Lagrangian numerical model, the expanding box model (EBM), first introduced by \citet{grappin1993PhRvL..70.2190G}, which allows us to decouple the internal MHD dynamics from the large-scale motion.
With the resulting high resolution, we investigate the impact of meso-scale internal processes on the evolution of the magnetic cloud; we focus on the interaction of the magnetic configuration with the anisotropic expansion and on the effect of superposing turbulent fluctuations to such a system.
Using EBM introduces an additional parameter to the usual MHD equations, the expansion rate, which controls the pace of spherical expansion and plays a key role in our study.

The structure of the paper is the following.
In Section~\ref{section-methods} we describe the expanding box model, the numerical setup, the magnetised equilibrium configuration used to describe the 2D section of a flux rope, and the main parameters used for the different simulations, focusing on the definition and meaning of the non-dimensional expansion rate $\eps_0$.
Sections~\ref{section-results_ideal} and~\ref{section-results_turb} present the results of our simulations of an expanding magnetic cloud, without and with turbulent fluctuations, respectively: the contributions of spherical expansion, internal dynamical balance and turbulence are qualitatively and quantitatively examined.
In Section~\ref{section-discussion} we compare our findings to previous theoretical and observational studies and examine our results in a broader context. The final section summarises our main findings and discusses limitations and perspectives of this work.
%
%
\section{Equations, numerical methods, initial conditions and parameters}
\label{section-methods}
\subsection{Ideal MHD equations in EBM}
In what follows we will employ the expanding box model (EBM, \cite{grappin1993PhRvL..70.2190G, grappin1996JGR...101..425G, rappazzo2005ApJ...633..474R, dong2014ApJ...793..118D}) to follow the evolution of a parcel of plasma in the spherically expanding solar wind.
Below we give a heuristic description of the EBM, focusing on its physical interpretation. A more rigorous derivation can be found in the above works.

We assume a mean wind velocity $\vec{U}_0 = U_0 \vvers_r$, which transports the parcel radially outward from the Sun. By doing so, the large-scale radial motion of the plasma is decoupled from the small-scale internal dynamics, which is numerically solved inside the simulation box.
The heliocentric position  of the parcel of plasma is given by
\begin{equation}
	R(t) = R_0 + U_0 t
\end{equation}
where $R$ is the radial (heliocentric) coordinate, starting at $R(t=0) = R_0$, and $U_0$ is the constant radial velocity.
Since we are considering a spherically expanding plasma, the box expands as a unit element of spherical volume: while moving outwards, the transverse size of the domain increases linearly with $R$ because of the assumed radial flow, while the radial size remains
constant. The expansion rate is given by $U_0 / R_0$, and is the inverse of the characteristic expansion time, $t_\mathrm{exp} = R_0 / U_0$, which is also the transport/propagation time.

In the following we will use the heliospheric distance normalised to its initial value:
\begin{equation}
	a = a(t) = \frac{R(t)}{R_0} = 1 + \frac{t}{t_\mathrm{exp}} .
\end{equation}%

The use of EBM implies a couple of relevant modifications to the usual MHD equations, which are now written in a reference frame that is comoving with the box.
In this frame, the large scale flow is absent and we are left with the velocity fluctuations with respect to the mean flow, $\vec{u} = \vec{U} - \vec{U_0}$. Now all variables depend only from the local coordinates,
$(x_\parallel^\prime = x_\parallel - U_0 t, x_{\perp_1}^\prime = x_{\perp_1}/a(t), x_{\perp_2}^\prime = x_{\perp_2}/a(t))$
where $\parallel$ and $\perp_1,~\perp_2$ refer to the radial direction and to the two transverse directions.
Spherical expansion implies that the transverse dimensions change with time (distance), while the radial one is constant,
\begin{equation}
\label{eqn-EBM-LprlLperp}
	L_\parallel = \mathrm{const.}(t) \quad ; \quad
	L_\perp \propto a(t)
	\quad\mathrm{for}\quad
	\perp = \perp_1, \perp_2
\end{equation}
The geometrical stretching is thus incorporated by multiplying the gradients in the usual MHD equations by anisotropic factors
\begin{equation}\label{eqn-EBM-scal-nabla}
	\nabla^\prime_\parallel = \nabla_\parallel \quad , \quad
	\nabla^\prime_\perp     = \frac{1}{a} \nabla_\perp
	\quad\mathrm{for}\quad
	\perp = \perp_1, \perp_2
\end{equation}
In what follows, we shall concisely write ``$\perp$'' without specifying ``for $\perp = \perp_1, \perp_2$''.

This brings to the non-dimensional ideal MHD equations in the EBM:
\begin{subequations}
\label{eqn-EBM}
\begin{eqnarray}
\frac{\partial\rho}{\partial t}
& = &
-\nabla'\cdot{\left( \rho\vec{u} \right)}
- 2\rho\frac{\dot{a}}{a}
\label{eqn-EBM-r}\\
\frac{\partial\vec{u}}{\partial t}
& = &
-\vec{u}\cdot\nabla' \vec{u}
-\frac{1}{\rho}\nabla'P
+\frac{1}{\rho}\left(\nabla'\times\vec{B}\right)\times\vec{B}
- \frac{\dot{a}}{a} \Pp \vec{u}
\label{eqn-EBM-u}\\
\frac{\partial\vec{B}}{\partial t}
& = &
\nabla'\times\left(\vec{u}\times\vec{B}\right)
+ \frac{\dot{a}}{a} \left(\Pp - 2\mathcal{I}\right) \vec{B}
\label{eqn-EBM-B}\\
\frac{\partial T}{\partial t}
& = &
-\vec{u}\cdot\nabla' T
- (\gamma-1) T \left(\nabla'\cdot\vec{u}\right)
- 2(\gamma-1)\frac{\dot{a}}{a} T
\label{eqn-EBM-T}
\end{eqnarray}
\end{subequations}
where thermodynamic variables are related through the equation of state $P=\rho T$ and all the variables are to be intended as functions of the comoving and expanding Cartesian coordinates $(x',y',z')$; moreover, $\mathcal{I}$ is the $3\times3$ identity matrix and $\Pp$ is the projector on the transversal directions
\begin{equation}\label{eqn-proj}
\left(\Pp\right)_{i,j} = \delta_{i,\perp}\delta_{j,\perp}
\delta_{i,j}
\qquad\textrm{for}\qquad
i,j = x', y', z' \quad .
\end{equation}
The frictional forces (i.e. the additional linear terms in the r.h.s. of Eqs.~\eqref{eqn-EBM}) make all the fields decay with different scaling laws:
for the number density
\begin{equation}\label{eqn-EBM-scal-n}
	n \propto a^{-2} \quad \text{which ensures mass conservation,}
\end{equation}
for the magnetic field components
\begin{equation}\label{eqn-EBM-scal-B}
	B_\parallel \propto a^{-2} \quad , \quad B_\perp \propto a^{-1} , \quad \text{(magnetic flux cons.)}
\end{equation}
and for the local velocity field components
\begin{equation}\label{eqn-EBM-scal-u}
	u_\parallel \propto \mathrm{const.} \quad , \quad u_\perp \propto a^{-1} \quad \text{(angular momentum cons.)}
\end{equation}
and finally the temperature decreases as
\begin{equation}\label{eqn-EBM-scal-T}
	T \propto a^{-2(\gamma-1)} \quad ,
\end{equation}
where $\gamma$ is the adiabatic index chosen equal to $5/3$.

The non-dimensional form is obtained defining the characteristic length ($L^0$), mass density ($\rho^0$), and magnetic field ($B^0$).
The other quantities are all expressed in Alfv{\'e}n units: velocities are measured in Alfv\'en speed, $u^0 = c_\mathrm{A}^0 = B^0/\sqrt{4\pi\rho^0}$; time in Alfv{\'e}n time, $t^0 = L^0 / u^0$; temperature in Alfv\'en units $T^0 = m_\mathrm{p} (u^0)^2 / 2 k_\mathrm{B}$.
Units are chosen relative to the initial configuration of the flux rope, so that $L^0 \equiv L_\mathrm{FR}$ and $B^0 \equiv B_\mathrm{FR}$ are the width and the axial field of the flux rope, while $\rho^0 \equiv \rho_\mathrm{bg}$ is the uniform ambient solar wind density (also equal to the flux rope density -- see next section for details on the initial configuration).

Since the expansion rate $U_0/R_0$ is an additional dimensional parameter with respect to the usual MHD equations, defined through $U_0$ and $R_0$, its non-dimensional counterpart is related to the velocity and length units (i.e. also to the magnetic field and density ones).
The non-dimensional expansion rate $\eps_0$ is then the ratio between the unit time $t^0$ (in our case, the characteristic Alfv{\'e}n time%
\footnote{%
In the previous works using EBM, the unit time was usually taken to be the non-linear time associated with turbulence. This definition will be recovered later.
}%
) and the expansion time $t_\mathrm{exp}$:
\begin{equation}\label{eqn-eps}
\eps_0 = \frac{t_\mathrm{A}}{t_\mathrm{exp}} = \frac{L_\mathrm{FR}}{R_0} \frac{U_0}{c_\mathrm{A}^0}
\quad ,
\end{equation}
and its value controls how fast the expansion is with respect to the flux rope's crossing time.
\subsection{Choice of geometry and numerical methods}
\begin{figure}[]
	\centering
	\resizebox{\hsize}{!}{\includegraphics{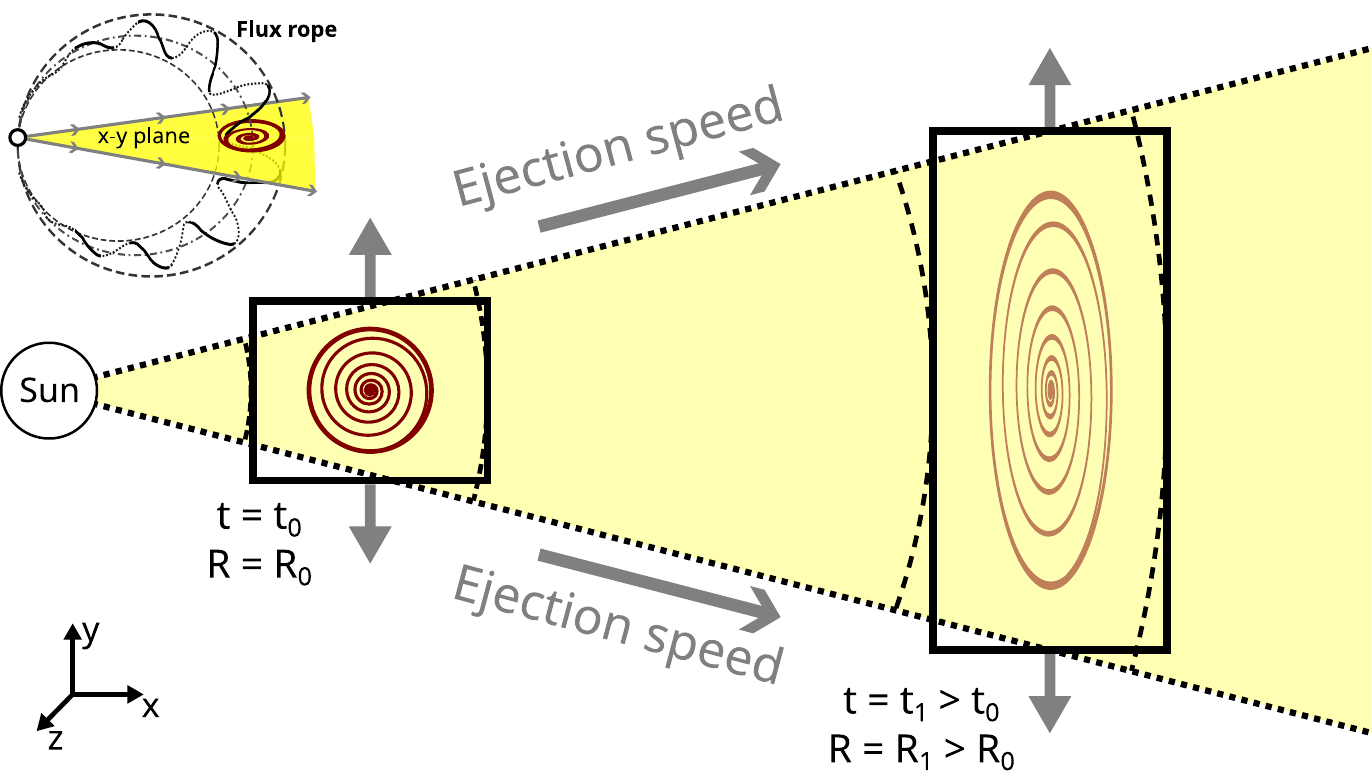}}
	\caption{
		Geometry and coordinate system for the expanding box model. We define $x$ as the spatial coordinate along the mean flow (local radial coordinate), $z$ as the local direction of the flux rope axis (along which the fields are assumed to be invariant), and $y$ completes the right-handed system. The flux rope has an initial circular section. The position of the box is $R(t)$, which increases with time as the flux rope propagates away from the Sun.
	}\label{fig-sketch-EBM}
\end{figure}

Throughout this work, we consider the coordinate $x_\parallel = x'$ as the locally radial coordinate and therefore $x_{\perp_1} = y'$ and $x_{\perp_2} = z'$ as the locally transverse coordinates. This implies that the projector on the transverse (non-radial) directions is $\Pp = \mathrm{diag}(0,1,1)$ and $\nabla' = (\partial/\partial_x, (1/a)\partial/\partial_y, (1/a)\partial/\partial_z)$.
We then consider a magnetic flux rope whose curvature radius is much larger than its initial radial size: this assumption of local cylindrical symmetry allows us to neglect the dynamics along the flux rope axis. We consider $z'$ to be directed along the axis, implying $\partial/\partial{z'} \equiv 0$, with the dynamics taking place in the plane $(x',y')$ normal to the axis. We also assume a 2.5D geometry, such that $\vec{u}$ and $\vec{B}$ have all three vector components. The 2D simulation domain is thus made of a radial ($x$) and a transversal/non-radial ($y$) direction.
A schematic representation of the configuration is shown in Fig.~\ref{fig-sketch-EBM}.
We are thus working in a comoving and expanding reference frame, containing the cross section of a magnetic cloud, which is moving away from the Sun with a constant radial velocity, meaning that we assume the plasma flow to be everywhere spherical (absence of non-radial motions) and uniform (no acceleration/deceleration of the large-scale flow); moreover, we assume the background solar wind to be non-magnetised.
In doing so, we implicitly impose that the internal kinematics of the flux rope follows the spherical expansion in absence of (internal or external) forces.

We assume all the fields to be periodic in the local coordinates $(x',y')$ and constant in the axial (out-of-plane) coordinate $z'$.
The numerical code integrates the viscous and resistive EBM equations on a domain $L_x \times L_y$, with a fixed grid $N_x \times N_y$, and is based on a pseudo-spectral 2.5D code already employed for example in \citet{papini2021ApJ...917L..12P}. Fourier transforms are used to compute spatial derivatives via the FFTW library \citep{frigo2005IEEEP..93..216F}, and a third order Runge-Kutta method is used for time advancement \citep{wray1990}.
The full equations are described in Appendix~\ref{appendix-equations}, and employ the out-of-plane component of the magnetic vector potential $A_z$ (in order to ensure $\nabla\cdot\vec{B} = 0$) and rescaled versions of the physical variables \citep[in order to use the equation for magnetic potential also with EBM, as devised by][]{rappazzo2005ApJ...633..474R}.

In addition to the physical variables, a passive scalar $s$ is evolved following an advection equation
\begin{equation}
	\label{eqn-ps}
	\partial_t s
	=
	-\vec{u}\cdot\nabla s
	\quad .
\end{equation}
The passive scalar is used as a tracer of the flux rope material and will be initialised in two regions (see next section).
\subsection{Initial equilibrium and flux rope identification}
\label{section_methods-init}
\begin{figure}[]
	\centering
	\resizebox{\hsize}{!}{\includegraphics{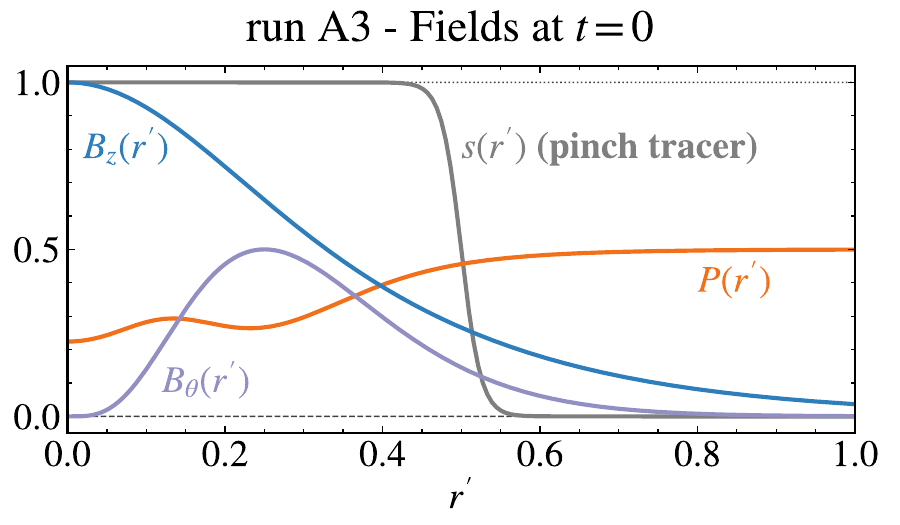}}
	\caption{
		Initial configuration of the main plasma parameters for the reference run {A3}: the solid lines show respectively the axial out-of-plane magnetic field $B_z$ (blue), poloidal in-plane magnetic field $B_\theta$ (violet), the kinetic pressure $P$ (orange) and the passive scalar $s$ as pinch tracer (grey), all as a function of the local radial coordinate $r'$.
	}\label{fig-fluxrope-init}
\end{figure}
We initialise the simulations by imposing a 2.5D profile for the flux rope equilibrium, using the local cylindrical coordinates $(r',\theta',z')$ to better exploit our assumption of cylindrical symmetry; the ($r'$-$\theta'$) plane corresponds to the $(x',y')$ plane normal to the flux rope axis.
No ambient magnetic field is present.

The configuration is static, stationary and axially symmetric around the flux rope axis. This means that $\vec{B} = B_{z'}(r') \vvers_{z'} + B_{\theta'}(r') \vvers_{\theta'}$ and $P = P(r')$.
With such assumptions the condition for a stationary and static equilibrium reads (from now on, the primes are dropped for better readability):
\begin{equation}
\label{eqn-init}
0
=
\frac{\partial}{\partial r} P(r)
+ \frac{1}{2} \frac{\partial}{\partial r} \left( {B_{\theta}(r)}^2 + {B_z(r)}^2 \right)
+ \frac{ {B_\theta(r)}^2 }{r}
\ .
\end{equation}
The equilibrium results from a balance of three forces: the magnetic pressure gradient ($B_\theta$ and $B_z$) that pushes away from the axis; the magnetic tension ($B_\theta$) and the kinetic pressure gradient ($P$) that pull towards the axis (the flux rope is colder than the environment).
We assume uniform density throughout the domain, equal to the background ambient solar wind, $\rho=\rho_\mathrm{bg}$; we choose a profile for both $B_\theta(r)$ and $B_z(r)$, thus leaving $P(r)$ to be analytically determined so as to account for the ultimate stability of the configuration.
The pressure profile is thus computed as
\begin{align}
P(r)
= &
- \frac{1}{2} \left( {B_\theta(r)}^2 + {B_z(r)}^2 \right)
- \int_0^r \frac{ {B_\theta(r^{\prime\prime})}^2 }{r^{\prime\prime}} \,\mathrm{d}r^{\prime\prime}
\\ \notag
&
+ \frac{1}{2} \left( {B_\theta(0)}^2 + {B_z(0)}^2 \right)
+ P(r=0)
\quad ,
\end{align}
where $P(r=0)$ is chosen so that $P_\infty = \rho_\mathrm{bg} T_\mathrm{bg}$.

The profiles of the in-plane poloidal magnetic field ($B_\theta$), the out-of-plane axial magnetic field ($B_z$), and pressure $P$ (which are described in Appendix~\ref{appendix-init}) are shown as a function of the local radial coordinate $r$ in Fig.~\ref{fig-fluxrope-init} with blue, violet and orange lines, respectively. The profile of the passive scalar $s$ is also shown as a grey line.
The magnetic field has an axial component which is zero at infinity and rises to a peak $B_{z,0} \equiv B_\mathrm{FR}$ ($1$ in code units) on the axis. The in-plane component vanishes at infinity and on the axis, with a peak $B_{\theta,0} = B_\mathrm{FR}/2$ ($0.5$ in code units) at $r = \Delta = L_\mathrm{FR}/4$ ($0.25$ in code units).

We initialise two passive scalars in two overlapping regions for later use in identifying the flux rope material, especially in turbulent simulations.
The ``pinch tracer'', $s_\mathrm{pinch}$, is initially limited to the region bounded by the in-plane (twisted) magnetic field.
As can be seen in the figure, the pinch tracer has a half-width of $L_\mathrm{FR}/2$ ($0.5$ in code units), but both $B_\theta$ and $B_z$ extend a bit further out because of their chosen functional form.
For this reason we also use also an ``axial tracer'', $s_\mathrm{axial}$, which is initialised to cover also the out-of-plane flux rope magnetic field, to keep track of its transport that is not constrained by the pinch provided by the in-plane magnetic field.
The primary use of the pinch tracer is to return an estimate of the radial and transversal sizes of the flux rope from the simulations. We use an integrated measure of the flux rope size, computed as the standard deviation of each coordinate using the pinch tracer $s$ as weight:
\begin{equation}
	\label{eqn-ps-sigma}
	\sigma_x = \left( \frac{ \iint (x - \bar{x})^2 \ s(x,y) \, \mathrm{d}{x} \, \mathrm{d}{y} }{ \iint s(x,y) \, \mathrm{d}{x} \, \mathrm{d}{y} }\right)^{1/2}
	\quad ,
\end{equation}
where
\[
\bar{x} = \frac{ \iint x \ s(x,y) \, \mathrm{d}{x} \, \mathrm{d}{y} }{ \iint s(x,y) \, \mathrm{d}{x} \, \mathrm{d}{y} }
\quad ,
\]
and similarly for $y$.

Once the simulation starts, expansion causes an anisotropic stretching of the domain and the decrease of field components and temperature, perturbing this initial equilibrium.
The anisotropic scaling of kinetic and magnetic pressure terms (coming from Equations~\eqref{eqn-EBM-scal-n}, \eqref{eqn-EBM-scal-B} and \eqref{eqn-EBM-scal-T}) already suggests a natural decrease of the plasma $\beta$, which would result in an over-expansion in all directions, since the kinetic pressure decays more rapidly with heliocentric distance than the magnetic pressure does, asymptotically:
\begin{equation}\label{eqn-beta}
\beta
\propto
\frac{\rho_{R_0} a^{-2} T_{R_0} a^{-4/3}}{B_{\perp,R_0}^2 a^{-2} + B_{\parallel,R_0}^2 a^{-4}}
\sim
\beta_0 a^{-4/3}
\quad .
\end{equation}
\subsection{Turbulent fluctuations and spectral quantities}
To study the effect of turbulence on the evolution of the flux rope in the expanding geometry, we perturb the initial equilibrium with pseudo-Alfv{\'e}nic fluctuations with zero divergence and zero mean cross helicity, $\delta\vec{u}$ and $\delta\vec{B}$, which naturally develop into MHD turbulence \citep[as in][]{papini2019ApJ...870...52P}.
We arbitrarily fix the amplitudes of the 2D Fourier coefficients $\delta\hat{B}$ and $\delta\hat{u}$ to decrease as $k^{-1/2}$, with $k=\sqrt{k_x^2 + k_y^2}$, resulting in an omnidirectional 1D flat (horizontal) power spectral density.
We further define $k_\mathrm{FR} = 2\pi / L_\mathrm{FR}$ and the minimum wave-number $k_0 = 2\pi/L_x$, the latter depending on the domain size.
Fluctuations are completely determined by the choice of two further parameters: the interval of excited (``turbulent'') scales, $k_\mathrm{turb}\in[k_{1},~k_{2}]$, and the root mean square (RMS) amplitude of fluctuations, $\delta B/\sqrt{4\pi\rho_\mathrm{bg}} \sim \delta{u}$.
Their value determines an additional non-dimensional parameter,
\begin{equation}\label{eqn-def-chi}
	\chi
	=
	\frac{ t_\mathrm{A} }{ t_\mathrm{NL} }
	=
	\frac{ 2\pi \, k_{1} \, \delta B }{ k_\mathrm{FR} \, B_{z,0} }
	\quad ,
\end{equation}
where $t_\mathrm{NL} = ( k_{1} \delta{u} )^{-1}$ is the non-linear time computed at the largest scale of turbulent fluctuations and where the Alfv{\'e}n time is the same as our unit time $t_\mathrm{A} = L^0 / c_\mathrm{A}^0 = L_\mathrm{FR} / (B_{z,0}/\sqrt{4\pi\rho_\mathrm{bg}})$.
Since we fixed $B_{\theta,0} / B_{z,0} = 1/2$ in our runs, the value of $\chi$ controls how fast the turbulent dynamics is compared to the flux rope's crossing time.
The ratio $\chi / \eps_0 = t_\mathrm{exp} / t_\mathrm{NL}$ defines the ``age'' of turbulence, that is, how fast turbulence evolves with respect to the expansion/propagation time; its inverse then defines the ``turbulent expansion rate'':
\begin{equation}\label{eqn-def-epsturb}
	\eps_\mathrm{T} = \eps_0 / \chi = t_\mathrm{NL} / t_\mathrm{exp}
	\quad .
\end{equation}
As a rule fluctuations are excited in the whole domain, including the flux rope, with one exception (see next section).

In analysing turbulent simulations we compute the power spectral density for the velocity and magnetic fluctuations.
The power spectral density of a vector field $\vec{f}$
is the omnidirectional 1D spectrum, obtained by summing the square of 2D Fourier coefficients, $\vec{\hat{f}}$, having wave-numbers that fall in a shell of thickness $[k_i,~k_{i+1})$,
\begin{equation}
	\label{eq-spectra-omni}
	\mathcal{P}_f(k_i) = \sum_{k_i\le|\vec{k}|<k_{i+1}}|\vec{\hat{f}}|^2
	\quad .
\end{equation}
The grid of shells of the omnidirectional spectrum is fixed at all times and equal to the wave-numbers of the non expanding direction, $k_i\in[k_x^{min},k_x^{max}]$; since the wave-numbers in the transverse direction decrease as $1/a$, we account for the anisotropic stretching by computing $|\vec{k}|=\sqrt{k_x^2+k_y^2/a^2}$.
With this choice the large wave-numbers in the transverse direction always contribute to the omnidirectional spectrum, but the large scales $k_y<k_x^{min}$ (including part of the flux rope) are left out.

As already pointed out by \citet{grappin1996JGR...101..425G} and \citet{dong2014ApJ...793..118D}, the expanding geometry weakens the spatial gradients and damps fluctuations, slowing down the non-linear dynamics of turbulence.
Recently \citet{pezzi2024A&A...686A.116P} perturbed a cylindrically symmetric flux rope by exciting 3D turbulence in the whole domain (homogeneous MHD, no spherical expansion); their results show that for weak fluctuations the large-scale coherent magnetic field inhibits the turbulent cascade towards small scales, whereas for strong fluctuations the cascade dominates and the flux rope can get strongly perturbed. Although in this study we are limited to 2D turbulence interacting with a 2.5D flux rope, we also aim to understand how much exciting fluctuations inside the flux rope counts in perturbing the equilibrium.
\subsection{Main parameters and simulations dataset}
\begin{table*}
	\caption{List of parameters for the simulations.}
	\label{table-runs}
	\centering
	\begin{tabular}{l c c c c c c c c c l}
		\hline\hline
        run & $\eps_0$\,\tablefootmark{a} & $\beta_0$\,\tablefootmark{b} & $\delta{B}$ & $k_\mathrm{turb}/k_\mathrm{FR}$ & $\chi$\,\tablefootmark{c} & $\eps_\mathrm{T}$\,\tablefootmark{d} &
        $\delta{B}_\mathrm{in/out}$ & $t_\mathrm{1AU}$ & $(L_x,L_y)\,[N_x \times N_y]$ & $\mu,\,\eta,\,\kappa$ \\
        \hline
        \texttt{A2}   & 0.2   & 1   &     &         &    &      &     & 32  & $(8,4)$ [$2048\times1024$] & \SI{7.5E-5}{} \\
        \texttt{A3}   & 0.4   & 1   &     &         &    &      &     & 16  & $(6,6)$ [$1024\times1024$] & \SI{7.5E-5}{} \\
        \texttt{A4}   & 0.8   & 1   &     &         &    &      &     & 8   & $(6,6)$ [$1024\times1024$] & \SI{7.5E-5}{} \\
        \texttt{A5}   & 1.6   & 1   &     &         &    &      &     & 4   & $(6,6)$ [$1024\times1024$] & \SI{1.0E-4}{} \\
        \texttt{A6}   & 3.2   & 1   &     &         &    &      &     & 2   & $(6,6)$ [$1024\times1024$] & \SI{2.0E-4}{} \\
		\hline
        \texttt{B3}   & 0.4   & 0.6 &     &         &    &      &     & 16  & $(6,6)$ [$1024\times1024$] & \SI{7.5E-5}{} \\
        \texttt{C3}   & 0.4   & 2   &     &         &    &      &     & 16  & $(6,6)$ [$1024\times1024$] & \SI{7.5E-5}{} \\
        \texttt{D3}   & 0.4   & 4   &     &         &    &      &     & 16  & $(6,6)$ [$1024\times1024$] & \SI{7.5E-5}{} \\
		\hline
        \texttt{Y3}   & 0.4   & 1   & 0.5 & $1{-}4$ &  3 & 0.12 & 1   & 16  & $(8,4)$ [$4096\times2048$] & $\SI{2.0E-4}{}\times{a^{-1}}$ \\
        \texttt{Y3k}  & 0.4   & 1   & 0.5 & $4{-}7$ & 13 & 0.03 & 1   & 16  & $(8,4)$ [$4096\times2048$] & $\SI{1.0E-4}{}\times{a^{-1}}$ \\
        \texttt{Y3h}  & 0     & 1   & 0.5 & $1{-}4$ &  3 &      & 1   & 16  & $(8,4)$ [$4096\times2048$] & $\SI{2.0E-4}{}$ \\
        \texttt{Z3}   & 0.4   & 1   & 0.5 & $1{-}4$ &  3 & 0.12 & 0.2 & 16  & $(8,4)$ [$4096\times2048$] & $\SI{2.0E-4}{}\times{a^{-1}}$ \\
        \hline
        \texttt{AR}   & 1.0   & 1   &     &         &    &      &     & 6.4 & $(6,6)$ [$1024\times1024$] & \SI{8.0E-5}{} \\
        \hline
	\end{tabular}
	\tablefoot{
		For all the simulations, $N_y$ is such that $\mathrm{d}y = L_y / N_y \equiv L_x / N_x = \mathrm{d}x$.
		\tablefoottext{a}{See Eq.~\eqref{eqn-eps}.}
		\tablefoottext{b}{See Eq.~\eqref{eqn-def-beta0}.}
		\tablefoottext{c}{See Eq.~\eqref{eqn-def-chi}.}
		\tablefoottext{d}{See Eq.~\eqref{eqn-def-epsturb}.}
	}
\end{table*}
We summarise now the main control parameters and comment briefly the simulations dataset used throughout the paper.
We recall that the initial equilibrium of the flux rope is the same in all runs and determines the code units, $B_{z,0} = 2 B_{\theta,0} = B_\mathrm{FR} = 1$, $L_\mathrm{FR} = 1$, $\rho_\mathrm{bg} = 1$.
\subsubsection{Expansion rate, plasma beta and fluctuations}
We carry out a series of simulations with a flux rope embedded in a uniform background (isolated flux rope, runs labelled with A, B, C, and D).
These runs are completely determined by two parameters:
the plasma $\beta$, common to standard MHD, here evaluated as the ratio between the external pressure and the flux rope's magnetic field
\begin{equation}\label{eqn-def-beta0}
	\beta_0 = 2 P_\mathrm{bg} / {B_\mathrm{FR}}^2
	\quad ,
\end{equation}
and the non-dimensional expansion rate $\eps_0$ (Eq.~\eqref{eqn-eps}), that controls how fast the spherical expansion is compared to the flux rope crossing time.
We will focus on run A3
to thoroughly describe the evolution and internal dynamics in Sections~\ref{section-results_ideal_evol} and~\ref{section-results_ideal_dynamics}.

The second series of simulations explores the impact of turbulence on the flux rope dynamics, using the same parameters as in run A3 and adding fluctuations (runs labelled with Y and Z); as a rule we excite fluctuations in the whole domain, thus also inside the flux rope (runs Y3, Y3h, Y3k), but since observations generally indicate that fluctuations are smaller inside the flux rope that outside of it \citep{regnault2020JGRA..12528150R,sorriso-valvo2021ApJ...919L..30S}, we also consider one case in which the amplitude of fluctuations is reduced by a factor $5$ when entering the flux rope (run Z3).
The addition of fluctuations introduces the non-dimensional parameter $\chi = t_\mathrm{A} / t_\mathrm{NL}$, which controls how fast the non-linear turbulent dynamics is compared to the flux rope crossing time; the turbulent expansion rate $\eps_\mathrm{T} = t_\mathrm{NL} / t_\mathrm{exp}$ then controls how fast the expansion is compared to the turbulence timescale.
We will focus on run Y3
to thoroughly describe the effects of turbulence in Sect.~\ref{section-results_turb}.

We do not attempt a complete scan of the parameter space, since it should also involve variation of the initial equilibrium (e.g. the degree of twist of the flux rope $B_{\theta,0}/B_{z,0}$) and of the turbulent spectrum (e.g. the range of excited wave vectors and its energy distribution); we prefer to select values that are reasonable in terms of the physical system of interest.

For the isolated runs, we choose values of $\eps_0 \in [0.2,3.2]$ (runs A2-A6), with more than an order of magnitude of separation.
Then we fix $\eps_0 = 0.4$ and vary instead $\beta_0$ by varying $T_\mathrm{bg}$, in a range $\beta_0 \in [0.6,4.0]$ (runs B3, A3, C3 and D3), with values below and above unity.

The simulations labelled with Y and Z have random fluctuations initially superposed to the flux rope: simulations Y3, Y3h and Y3k all have the same parameters as A3 and $\delta{B} / B_\mathrm{FR} = 0.5$: run Y3h is the same as Y3 but with no expansion (regular MHD); run Y3k is the same as Y3 but with fluctuations at smaller scales; run Z3 is the same as Y3 but with smaller fluctuations inside the flux rope.

For the turbulent runs, our choices of $\delta B / B_{\theta,0} \sim 1$ (strong turbulence) and $k_\mathrm{turb} / k_\mathrm{FR} \gtrsim 1$ (fluctuations on smaller scales than the flux rope) imply $\chi \gtrsim 3$, that is, a quickly developing turbulence with respect to the flux rope crossing time.
The expanding runs Y3, Y3k and Z3 all have $\eps_0 = 0.4$, so $\eps_\mathrm{T} \lesssim 0.13$ (fast turbulence compared to expansion); a value $\eps_\mathrm{T} \lesssim 1$ is required to have a dynamically relevant turbulence during the propagation, otherwise the non-linear cascade would take too much time and the turbulence would be ``frozen'' during the propagation. Run Y3k has an $\eps_\mathrm{T}$ four times smaller than run Y3, that is, a quicker/shorter-lived turbulence.

For runs labelled with A, B, C, and D, we use open boundaries in the $x$ direction \citep[similarly to][]{rappazzo2005ApJ...633..474R}, whose details will be given in a separate work. The results were tested to be the same with open and periodic boundaries, but using open boundaries allows us to avoid spurious reflections without needing a wider domain and more grid points.
\subsubsection{Interpretation of the non-dimensional expansion rate and limitations}
\label{section-methods_eps0}
Since the non-dimensional expansion rate $\eps_0$ is given by four different quantities (recall Eq.\eqref{eqn-eps}), a value of $1$ corresponds equally to a flux rope starting at $R_0 = 5 L_\mathrm{FR}$ with a speed $U_0 = 5 c_\mathrm{A}^0$, or to a flux rope that is initially closer to the Sun and moves slower of the same factor.
However, our assumption of constant speed is not well justified close to the Sun, because of the initial driving phase to which CMEs are subject; a CME can be thought of as isolated and interacting with just the solar wind only after a distance of about $20$ to $30{R_\sun}$ \citep[see e.g.][]{st-cyr2000JGR...10518169S, subramanian2007A&A...467..685S, sachdeva2015ApJ...809..158S}.

We thus fix the starting heliocentric distance to be $30 R_\sun$ for all the simulations.

Our parameter $\eps_0$ then reflects three free parameters in $L^0 = L_\mathrm{FR}$, $U_0$ and $c_\mathrm{A}^0 = B^0 / \sqrt{4\pi\rho^0} = B_\mathrm{FR} / \sqrt{4\pi\rho_\mathrm{bg}}$, which all contribute to set the ratio between expansion and Alfv{\'e}n timescales. However, CMEs have a quite diverse range of sizes, velocities and magnetic field intensities, possibly resulting in a broad range of values for $\eps_0$ because of a difference in one quantity, or equal values of $\eps_0$ for very different CMEs.
To ease the physical interpretation of $\eps_0$, one can think of fixing the background density and magnetic cloud peak field, thus fixing $c_\mathrm{A}^0$, and also the initial flux rope size: varying $\eps_0$ then corresponds to varying the CME propagation velocity.
The relation of $\eps_0$ to actual flux rope parameters will be further discussed in Sect.~\ref{section-discussion}, and run AR will also be described.
It should be kept in mind that the meaning of $\eps_0$ is the ratio between the internal Alfv{\'e}n and the large-scale propagation/expansion timescales.

Because of the functional form of our initial equilibrium configuration, which requires a dip in the temperature profile in order to confine the magnetic flux rope, we cannot choose a background temperature too low, since the resulting value inside the flux rope would become negative.
This possibly limits us to a narrower range of $\beta_0$ values than what is generally observed (again, see discussion in Sect.~\ref{section-discussion}).

Independently of the other parameters, all simulations are run up to the same final distance, $R = 7.4 R_0 \simeq \SI{1}{AU}$, that is: the smaller the expansion rate, the longer the time it takes to get to $\SI{1}{AU}$ (see $t_\mathrm{1AU}$ in Table~\ref{table-runs}).
%
%
\section{Results: Internal dynamics, spherical and radial expansion}
\label{section-results_ideal}
\begin{figure*}[]
	\centering
	\includegraphics[width=17cm]{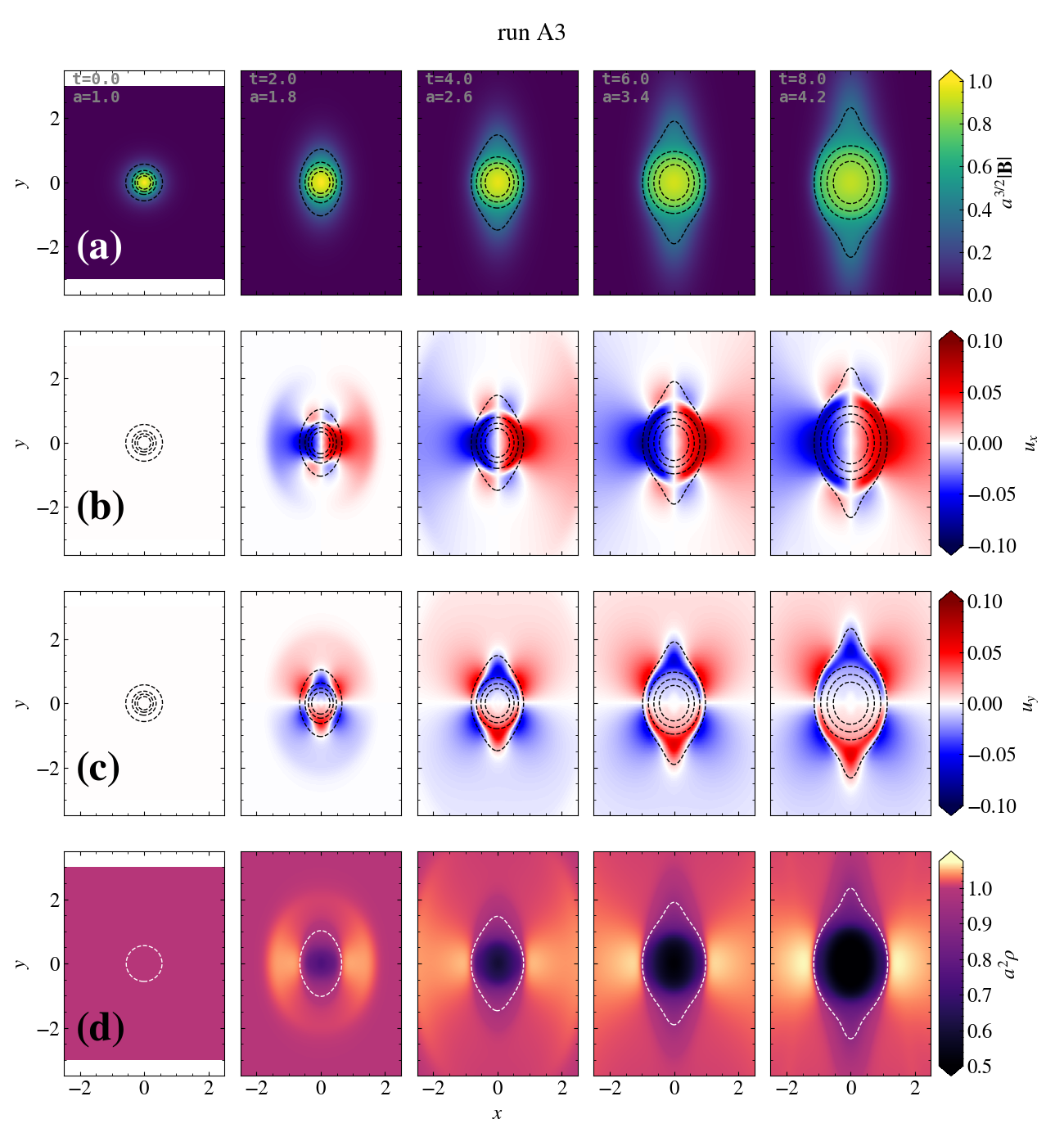}
	\caption{
		Run {A3}: from top to bottom, evolution of magnetic field $|\mathbf{B}|$ (a), velocity $u_x$ (b) and $u_y$ (c) and density $\rho$ (d). For  $|\mathbf{B}|$ and $\rho$ the fields' decay with heliocentric distance has been compensated for better visualisation.
		The in-plane magnetic field is represented in panels (a-c) as constant-$A_z$ black lines, whereas the outermost boundary of the pinch tracer is represented in panel (d) as a white dashed line.
		All the quantities are in code units.
	}\label{fig-maps}
\end{figure*}
\begin{figure*}[]
	\centering
	\includegraphics[width=17cm]{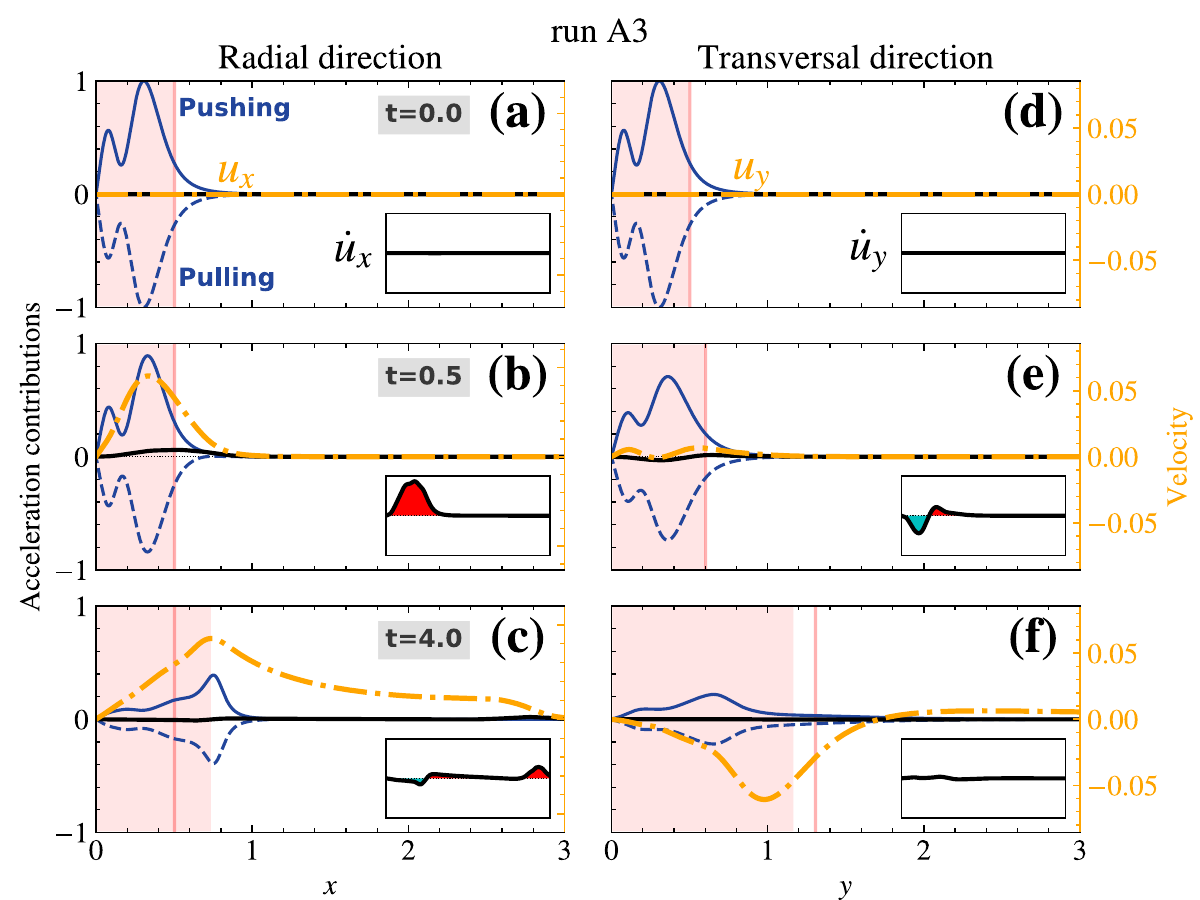}
	\caption{
		Run {A3}: evolution of the main dynamical terms (solid and dashed blue lines), their resultant (black line) and the velocity field (orange dash-dotted line, different scale), along both $x$ (left panels) and $y$ (right panels). The 1D cuts are taken at half domain (i.e. they intersect the flux rope axis). All the quantities are in code units and the dynamical terms are rescaled relative to their maximum at $t=0$. Only the right half of each 1D spatial domain is shown, since the dynamics is just inverted in the other half (axial symmetry). The computed (expected) pinch tracer size is shown as a light pink filled region (vertical line) for each snapshot.
		In the bottom right corner of each panel, an inset shows just the resultant acceleration, $\dot{u}_{x,y}$, as a black line filled with red/cyan according to its positive/negative sign.
	}\label{fig-acceleration-terms}
\end{figure*}
\begin{figure*}[]
	\centering
	\includegraphics[width=17cm]{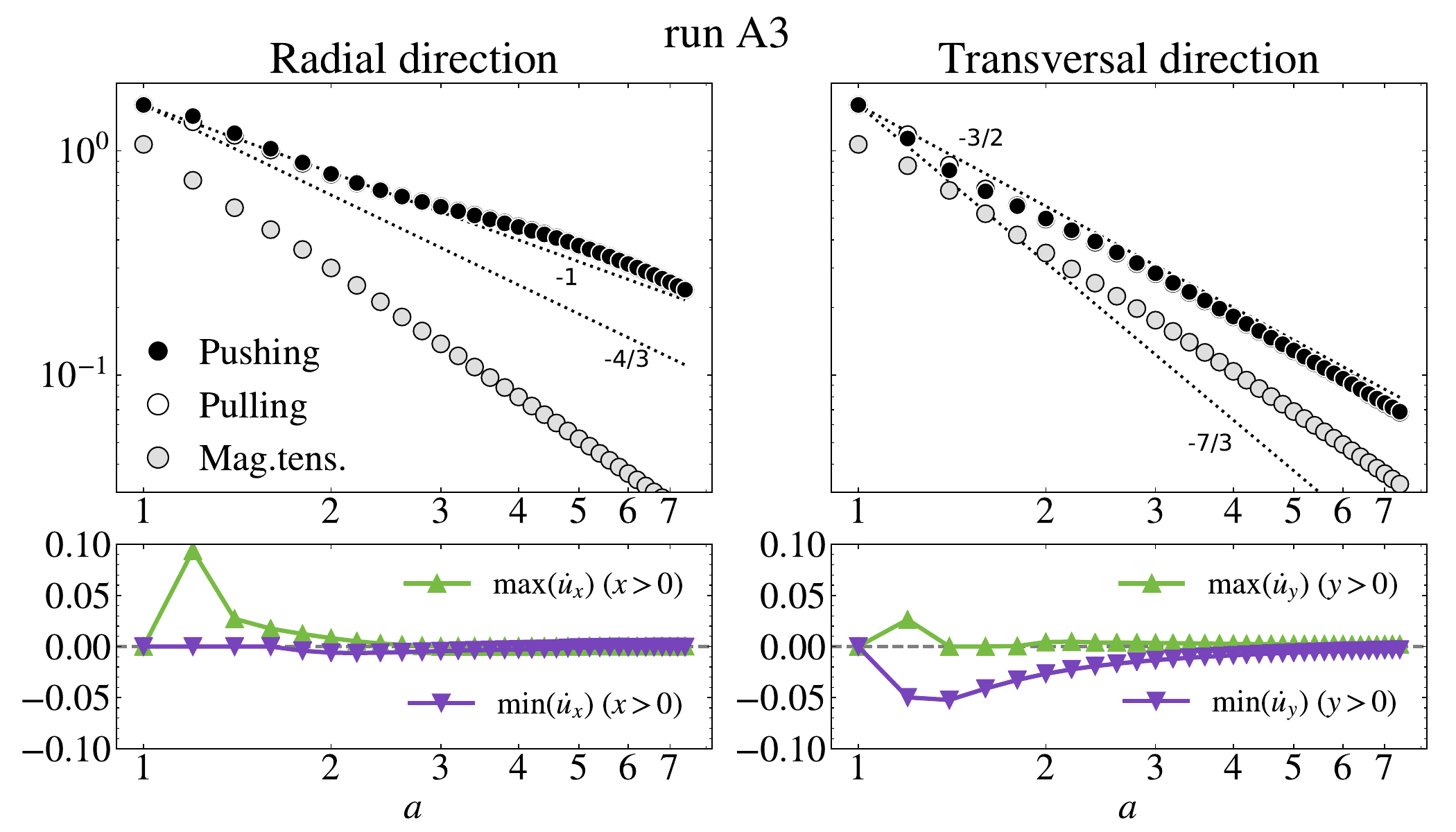}
	\caption{
		Run {A3}: evolution of the peaks of the dynamical terms with heliocentric distance, along $x$ (left) and $y$ (right). The peaks are computed inside the region of non-zero pinch tracer, in order to filter out the outward propagating shock front. The top panels show the absolute value of the main terms: pushing and pulling terms are drawn with black and white filled circles, and magnetic tension with light grey circles; black dotted lines represent the predicted scaling laws for magnetic and kinetic pressure gradient terms as shown in  Eq.~\eqref{eqn-xy-stability}. The bottom panels show the maximum and minimum of the resultant, in green and purple respectively. All the terms are in code units.
	}\label{fig-acceleration-scalings}
\end{figure*}
\subsection{Heliospheric evolution}
\label{section-results_ideal_evol}
The evolution of magnetic field ($|\vvec{B}|$), radial velocity ($u_x$), transverse velocity ($u_y$), and density ($\rho$) is shown in Fig.~\ref{fig-maps} for run {A3}. Time (or distance) runs from left to right in each row and is indicated in the top panels.
We note that for a better visualisation, only a part of the transverse domain is shown and all quantities are rescaled by the average decay expected from conservation of linear invariants, Eqs.~\eqref{eqn-EBM-scal-n}-\eqref{eqn-EBM-scal-u}, as indicated next to the colour bars.

The in-plane magnetic field is also shown as constant-$A_z$ solid lines in Fig.~\ref{fig-maps}(a-c), whereas the outermost boundary of the pinch tracer is shown in Fig.~\ref{fig-maps}(d) as a dashed line.
The tracer follows pretty well the flux rope boundary as identified by the in-plane magnetic field and is redundant at this stage.
However, it will be necessary to identify the flux rope material in presence of turbulent fluctuations in Sect.~\ref{section-results_turb}. For consistency, we  will use the tracer throughout the paper to determine the flux rope extent along the $x$ and $y$ directions.

In the top panels one can see that during the radial propagation, the isotropic initial equilibrium configuration is stretched in the transverse direction due to the large-scale motion of the box, as expected from the spherical geometry; a closer view of panel (a) shows that the flux rope's radial size $S_x$ increases with time, whereas its transverse size $S_y$ seems to increase less than linearly with distance; the result is an anisotropic configuration.
We recall that a passive plasma structure subject to spherical expansion would undergo a ``kinematic expansion'', that is, its radial size would remain constant whereas its transversal size would increase linearly with heliocentric distance (recall Eq.~\eqref{eqn-EBM-LprlLperp}).
The flux rope's radial expansion implies an aspect ratio smaller than the ``kinematic'' one, at any given distance.

Deviations from the kinematic expansion can be roughly explained by considering the evolution of the total dynamical pressure.
Since the magnetic field is present only inside the flux rope and recalling Eq.~\eqref{eqn-beta}, it follows that
\begin{equation}
	\frac{P^\mathrm{TOT}_\mathrm{in}}{P^\mathrm{TOT}_\mathrm{out}}
	\simeq
	\frac{(\rho T)_\mathrm{in} + B^2/2}{(\rho T)_\mathrm{out}}
	\simeq
	1 + \beta^{-1}
	\sim
	1 + C a^{4/3}
	\quad ,
\end{equation}
so that the internal over-pressure produces a (locally) outward force, which brings to the radial expansion.

Inspection of the top panels in Fig.~\ref{fig-maps} shows the out-of-plane $B_z$ (colour-coded) and the in-plane $B_\theta$ (black iso-lines) magnetic field components to follow slightly different evolutions: along the radial direction the extent of both components is about the same, but in the transverse direction $B_z$ extends more than $B_\theta$.
Also, the velocity components $u_x$ and $u_y$ in panels (b) and (c), respectively, and the density in panel (d), reveal a richer dynamics.

As soon as expansion is triggered, a pressure wave is launched and propagates anisotropically away from the flux rope.
This wave compresses the surrounding plasma and causes a dramatic drop of density inside the flux rope; the temperature $T$ -- not shown -- has a very similar evolution, with heating outside and cooling inside; this perturbation contributes to keep $\beta<1$ inside the flux rope.
At the same time a radial velocity pushes the magnetic field away from the flux rope centre, while the transverse velocity pulls it towards the centre.
The larger transverse extent of $B_z$ compared to $B_\theta$ is the result of this pulling force and of the initial condition ($B_z$ tail extending beyond the region where $B_\theta{\ne}0$), and is not due to magnetic diffusion (we will come back to this point later).
\subsection{Dynamical analysis}
\label{section-results_ideal_dynamics}
A more quantitative description of the internal dynamics and kinematics of the flux rope can be obtained by looking at Fig.~\ref{fig-acceleration-terms}, where the extent of the flux rope tracer is shown as a shaded pink along 1D cuts passing through the centre in the $x$ and $y$ directions (left and right panels, respectively), for different times, $t=0, 0.5, 4.0$ (from top to bottom); only half of the 1D domain is shown because it is symmetric with respect to the flux rope centre.
The theoretical flux rope extent as due to the kinematic stretching of the box is drawn with a dark pink vertical line.

By comparing the two bottom panels, it is clear that along the radial direction the flux rope has expanded, contrary to the constant-size prediction by simple kinematic expansion, while in the transverse direction the flux rope has expanded less than kinematically.

To describe the dynamics of the flux rope, we also show in each panel the profile of the pushing and pulling acceleration terms.
Pushing forces point away from the flux rope axis and are given by the magnetic pressure term (mainly the out-of-plane component). Their resulting acceleration is plotted with solid blue lines.
Pulling forces point towards the flux rope axis and are given by magnetic tension (involving only $B_\theta$) and the external (gas) pressure. Their resulting acceleration is plotted with dashed blue lines.
The solid black line is the overall resultant acceleration, including all the terms in Eq.~\eqref{eqn-EBM-u}); the dashed orange line is the profile of the velocity along the cut plotted on a different scale for better visualisation ($u_x$ and $u_y$ in the left and right panels, respectively); the inset shows again the total acceleration for better visualisation.

Along the radial direction ($x$), a magnetic over-pressure produces a pushing force at the edge of the flux rope, which in turn creates the outward-directed velocity, that is, the radial expansion.
Along the transversal direction ($y$), a pushing force is present as well but weaker, and the main dynamical term becomes magnetic tension, which produces a pulling force located inside the flux rope, which creates an inward directed velocity, that is, a local contraction.

The wavefront already visible in Fig.~\ref{fig-maps} can be more clearly identified, especially along $x$ (centre and bottom rows); also, the prevailing of velocity-related dynamical terms such as $-\vec{u}\cdot\nabla\vec{u}$ at later times can be noted (bottom row); this information is especially visible in the inset of Fig.~\ref{fig-acceleration-terms}(c).
A hint to the different evolution of $B_z$ and $B_\theta$ along $y$ is visible in Fig.~\ref{fig-acceleration-terms}(e), where positive and negative forces coexist around the boundary of the flux rope.

The forces involved in the flux rope equilibrium can be further analysed: the relative importance of the forces at early times can be estimated using the scaling laws for the evolution of gradients and fields in the EBM, namely Eq.~\eqref{eqn-EBM-scal-nabla} and Eqs.~(\ref{eqn-EBM-scal-n},~\ref{eqn-EBM-scal-B},~\ref{eqn-EBM-scal-T}) (along with $\gamma=5/3$). Substituting those scaling laws in Eq.~\eqref{eqn-EBM-u} and explicitly considering the $x$ and $y$ components, we obtain
\begin{multline}\label{eqn-xy-stability}
\frac{\mathrm{d}u_x}{\mathrm{d} t}
\propto
-\frac{1}{\rho_{R_0}}\left[
a^{-4/3}|\nabla_x P|_{R_0} +a^{-1}|\nabla_x P_\mathrm{m}|_{R_0} - a^{-2}|\mathbb{T}_x|_{R_0}\right]
\\
\frac{\mathrm{d}u_y}{\mathrm{d} t}
\propto
-\frac{1}{\rho_{R_0}} \left[ a^{-7/3}|\nabla_y P|_{R_0}   + a^{-3/2}|\nabla_y P_\mathrm{m}|_{R_0} - a^{-1}|\mathbb{T}_y|_{R_0}\right]
\ \ ,
\\
\end{multline}
where on the r.h.s the three terms represent the predicted scaling laws of kinetic and magnetic pressure gradients (respectively $P$ and $P_\mathrm{m}$) and magnetic tension ($\mathbb{T}$). Since the magnetic field components $B_\parallel$ and $B_\perp$ follow different scaling laws, for brevity we used their geometrical mean to evaluate the overall magnetic pressure gradient scaling.
Along $x$, the magnetic pressure gradient dominates, explaining the net pushing force that is responsible for the increase of the radial size. The magnetic tension has the fastest decrease, suggesting a later dynamics dominated by magnetic and gas pressures.
On the contrary, along $y$ the magnetic tension dominates, explaining the net pulling force responsible for the less-than-kinematic increase of the transverse size. In this direction the gas pressure strongly decreases, suggesting a purely magnetic kind of dynamics between magnetic pressure and tension.

To evaluate the relative importance of the forces also at later times for run {A3}, in Fig.~\ref{fig-acceleration-scalings} we plot the heliocentric evolution of the pushing and pulling terms (top panels), separately for the radial (left) and transverse direction (right). In the top panels the expected scaling laws from Eq.~\eqref{eqn-xy-stability} are also shown with dotted lines, while the contribution of the magnetic tension is plotted with filled grey circles.
At any distance we compute quantities inside the region traced by the pinch tracer in order to filter out the outward propagating acceleration associated with the wave. The difference between the pushing and pulling peak values returns a good estimate of the net acceleration. However, the maximum and minimum are not always co-spatial, so both positive and negative acceleration regions can coexist. This further information is supplied in the bottom panels where the minimum and maximum of the total resultant acceleration are plotted.
The pushing term is the absolute value of the extremum of the pushing acceleration profile inside the flux rope (cf. solid blue line inside the shaded area in Fig.~\ref{fig-acceleration-terms}); the other terms are computed in the same way%
\footnote{%
	Other dynamical terms arise after the first dynamical phase, in addition to the ones considered so far: mainly the advection $-\vec{u}\cdot\nabla\vec{u}$ and, along $y$, the EBM-related frictional deceleration $-\mathcal{P}_\perp (\dot{a}/a) \vec{u}$. Therefore, the total acceleration is not everywhere zero (cf. the inset in panel (c) of Fig.~\ref{fig-acceleration-terms}).
}%
.

Along the radial direction (left panels): for short distances the previous scaling analysis based on Eq.~\eqref{eqn-xy-stability} seems to hold: magnetic tension is negligible, magnetic pressure dominates on the gas pressure, giving an outward net acceleration which produces the radial expansion; at the same time, this same dynamical imbalance triggers a compression-rarefaction wave which enhances the gas pressure gradient, soon restoring the equilibrium which lasts for successive times.

Along the transversal direction (right panels): the initial dynamics is a bit more complex, with positive and negative accelerations (bottom panel) that cause a non homogeneous deformation of the flux rope. The dominant acceleration is the inward one, producing a less-than-kinematic transversal expansion. The magnetic tension contributes almost equally to the gas pressure in the pulling term at all distances. Eventually an equilibrium is reached as well.
\subsection{Summary}
The overall picture for a magnetic flux rope in a spherically expanding flow (exemplified with run {A3}) can then be summarised as follows:
\begin{enumerate}
\item the initial static and stationary equilibrium is perturbed by the anisotropic spherical expansion as the box is set in motion; the resulting dynamical imbalance produces accelerations in both directions: along the radial direction the flux rope is pushed away from its axis, whereas along the transversal direction it is pulled towards it;
\item such forces produce local velocities which act on the plasma, counteracting the geometrical stretching and causing a radial expansion and a transversal contraction (or less than geometrical expansion);
\item after the initial dynamical phase, the system reaches an equilibrium configuration, with the local velocities still present: the radial expansion and transversal contraction remain while the box is being transported away from the Sun by the large-scale flow, and evolves through a series of successive equilibria.
\end{enumerate}
%
\subsection{Dependence of flux rope size on the expansion rate and plasma $\beta$}
\label{section-results_ideal_parameters}
\begin{figure}[]
	\centering
	\resizebox{\hsize}{!}{\includegraphics{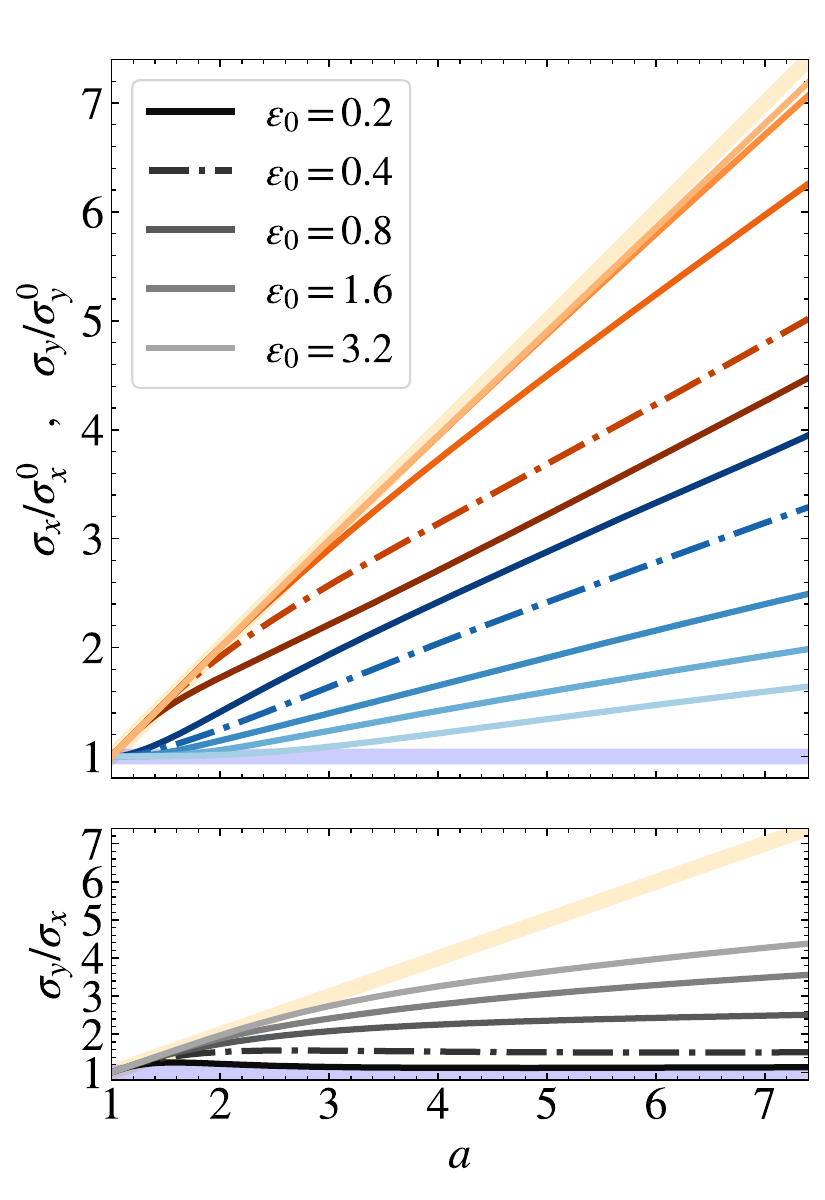}}
	\caption{
		Heliospheric evolution of the flux rope size estimates $\sigma_x$ and $\sigma_y$ (top panel) and the $y$-$x$ aspect ratio (bottom panel) for runs A2, A3, A4, A5, A6, that is, increasing $\eps_0$. Lighter shades represent increasing parameter values. Run {A3} is represented as a dash-dotted line. The light blue (orange) broad line represents the expected ``kinematic'' trend along $x$ ($y$).
	}\label{fig-sigmas-eps0}
\end{figure}
\begin{figure}[]
	\centering
	\resizebox{\hsize}{!}{\includegraphics{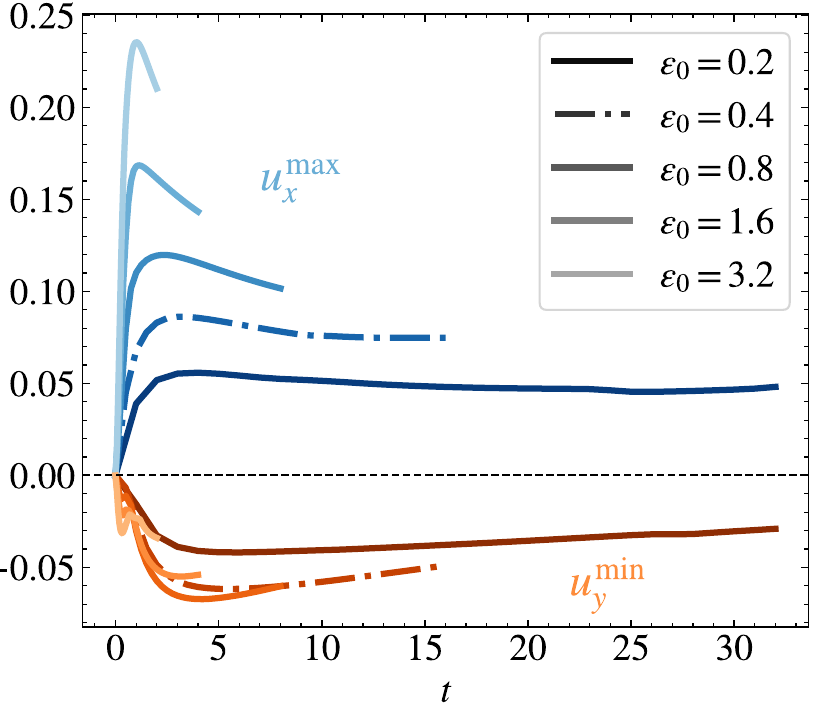}}
	\caption{
		Time evolution of the maximum of $u_x$ and minimum of $u_y$ (both computed across a 1D cut passing through the axis, as in Fig.~\ref{fig-acceleration-terms}) for runs A2-A6. The trends of $u_x^\mathrm{max}$ and $u_y^\mathrm{min}$ are drawn respectively with blue and orange lines, with lighter shades corresponding to higher values of $\eps_0$ as in Fig.~\ref{fig-sigmas-eps0}.
	}\label{fig-maxvxvy-t-ee0}
\end{figure}
We now study how the radial and transverse sizes vary with the two non-dimensional parameters, the non-dimensional expansion rate $\eps_0$ and the effective plasma beta $\beta_0$.

We first focus on the expansion rate.
On the one hand the anisotropic spherical expansion perturbs the equilibrium on expansion timescales $\tau_\mathrm{exp}$. On the other hand, the flux rope reacts against perturbations on a timescale related to the in-plane Alfv{\'e}n time $\tau_\mathrm{A}$.
The non-dimensional expansion rate $\eps_0$ thus controls the relative importance of the continuous geometrical stretching (and its consequences on the decrease of gradients and fields) and of magnetic tension reactivity.

In Fig.~\ref{fig-sigmas-eps0} we plot flux rope size estimates, $\sigma_x$ and $\sigma_y$, as a function of the heliocentric distance for runs A2-A6, that is, for increasing expansion rate $\eps_0$. The values of $\sigma_{x,y}$ are rescaled to their initial value for better readability; $\sigma_x$ and $\sigma_y$ are coloured in shades of blue and orange respectively, with lighter shades representing higher expansion rates $\eps_0$; the trends expected from simple kinematic expansion are drawn as thick light lines of the same colours; run {A3} discussed in the previous section is represented as a dash-dotted line.
For higher expansion rates, at a given distance the transverse size gets larger, approaching the kinematic behaviour, whereas the radial size gets smaller but keeps its increasing trend even for the strongest expansion rate considered here.

Independently of the expansion rate, the trends of $\sigma_{x,y}$ exhibit two stages: a first transient with variable slope, and a more stable phase with a more constant slope. This is consistent with the dynamical analysis shown in Fig.~\ref{fig-acceleration-scalings}, with a first phase of dynamical imbalance and a second phase of expanding equilibrium; moreover, the transient lasts longer along $y$ than along $x$, consistently with the persistence of net accelerations along the same directions. Finally, for higher expansion rates the variable transient lasts for a longer distance.

The aspect ratio $\sigma_y/\sigma_x$ is also plotted versus distance in the bottom panel of Fig.~\ref{fig-sigmas-eps0}, using shades of grey with the same meaning as in the top panel.
A critical expansion rate $\eps_0^*$ can be visually identified such that for $\eps_0\lesssim\eps_0^*$ (runs A2-A3) the aspect ratio attains an asymptotic value close to one (both directions continue to expand but at the same rate) whereas for $\eps_0\gtrsim\eps_0^*$ (runs A4-A6) the aspect ratio continues to increase without reaching a constant value.
Even for the highest expansion rate considered here, the kinematic trend of the aspect ratio is not reached because of the radial expansion.

The critical expansion rate can be estimated considering the 2D magnetic tension communication timescale between $x$ and $y$: the azimuthal Alfv{\'e}n speed is $c_{\mathrm{A},\theta} = (B_{\theta,0}/B_{z,0}) c_\mathrm{A}^0$, and it has to travel the length of a quarter-circle arc $L_\theta = \pi L_\mathrm{FR} / 4$; this implies that the effective expansion rate is
\begin{equation}
	\label{eqn-eps0-eff}
	\eps_{0,\mathrm{eff}} = \tau_{\mathrm{A},\theta} / \tau_\mathrm{exp} \sim (\pi/4) (B_{z,0}/B_{\theta,0}) \eps_0
	\quad ;
\end{equation}
then $\eps_{0,\mathrm{eff}} \lessgtr 1$ corresponds to $\eps_0 \lessgtr (4/\pi) (B_{\theta,0}/B_{z,0})$ and with our choice of $B_{\theta,0}/B_{z,0} = 1/2$ we have $\eps_0^* = 2/\pi \simeq 0.637$.

The different behaviour for runs with $\eps_0 \lessgtr \eps_0^*$ can be understood by considering how the expansion rate evolves with distance.
Following Eqs~\eqref{eqn-EBM-LprlLperp},~\eqref{eqn-EBM-scal-n} and~\eqref{eqn-EBM-scal-B}, the Alfv{\'e}n timescales along the $x$ and $y$ direction both scale%
\footnote{%
	Actually, the effective timescale for the $2D$ communication mediated by magnetic tension would be better estimated as the time it takes for an Alfv{\'e}n wave to travel along the elliptical arc of angular width $\pi/2$, varying minor and major axes $L_{\mathrm{FR},x}$ and $L_{\mathrm{FR},y}$, with varying speeds $c_{\mathrm{A},x}$ and $c_{\mathrm{A},y}$. It can however be approximated as $\tau_\mathrm{A} \propto a$.
}
as $\tau_\mathrm{A} \propto a$.
The expansion timescale at a given distance can be estimated from Eq.~\eqref{eqn-EBM-u} and also scales as $\tau_\mathrm{exp} \sim a/\dot{a} = a/\eps_0$.
Thus the non-dimensional expansion rate at any distance stays constant and only depends of its initial value:
\begin{equation}
	\eps(R) = \tau_\mathrm{A}(R) / \tau_\mathrm{exp}(R) \equiv \eps_0
	\quad .
\end{equation}
This in turn implies that if $\eps_0$ is initially set greater (smaller) than $\eps_0^*$, then the flux rope will remain causally connected (or disconnected in some regions) during the expansion. In addition, simulations with different values of $\eps_0$ are not just shorter or longer portions of the same evolution, but intrinsically different ones.

By increasing the non-dimensional expansion rate, we are considering magnetic clouds that move (and are anisotropically stretched) faster and faster with respect to the timescales of their internal processes (e.g. Alfv{\'e}n waves); a higher expansion rate implies a faster stretching, but also a stronger dynamical imbalance and thus more intense velocity fields; however, a faster magnetic cloud has also less time for its internal dynamics to develop, that is, for the internal velocities to act on the flux rope.
In Fig.~\ref{fig-maxvxvy-t-ee0} we show the evolution of the maximum radial and minimum transversal velocities ($u_x^\mathrm{max}$ and $u_y^\mathrm{min}$) with time for different expansion rates; $u_x^\mathrm{max}$ and $u_y^\mathrm{min}$ are directly related to radial expansion and to resistance to transversal expansion, respectively; the graphical conventions are the same as for Fig.~\ref{fig-sigmas-eps0}.
As the expansion rate increases (lighter lines), $u_x^\mathrm{max}$ gets higher, but this increase is less than linear with $\eps_0$ (scales approximately as ${\eps_0}^{1/2}$); since the travel time for fixed initial and final distances scales as ${\eps_0}^{-1}$ by definition, the result is a radial expansion which decreases for higher expansion rates.
For the transversal velocity, the same reasoning applies, but the EBM frictional force further damps $|u_y|$, resulting in weaker transversal resistance for higher expansion rates.
Overall, these two effects imply that higher $\eps_0$ correspond to smaller deviations from the predicted kinematic evolution and a more elongated aspect ratio, explaining Fig.~\ref{fig-sigmas-eps0}.

\begin{figure}[]
	\centering
	\resizebox{\hsize}{!}{\includegraphics{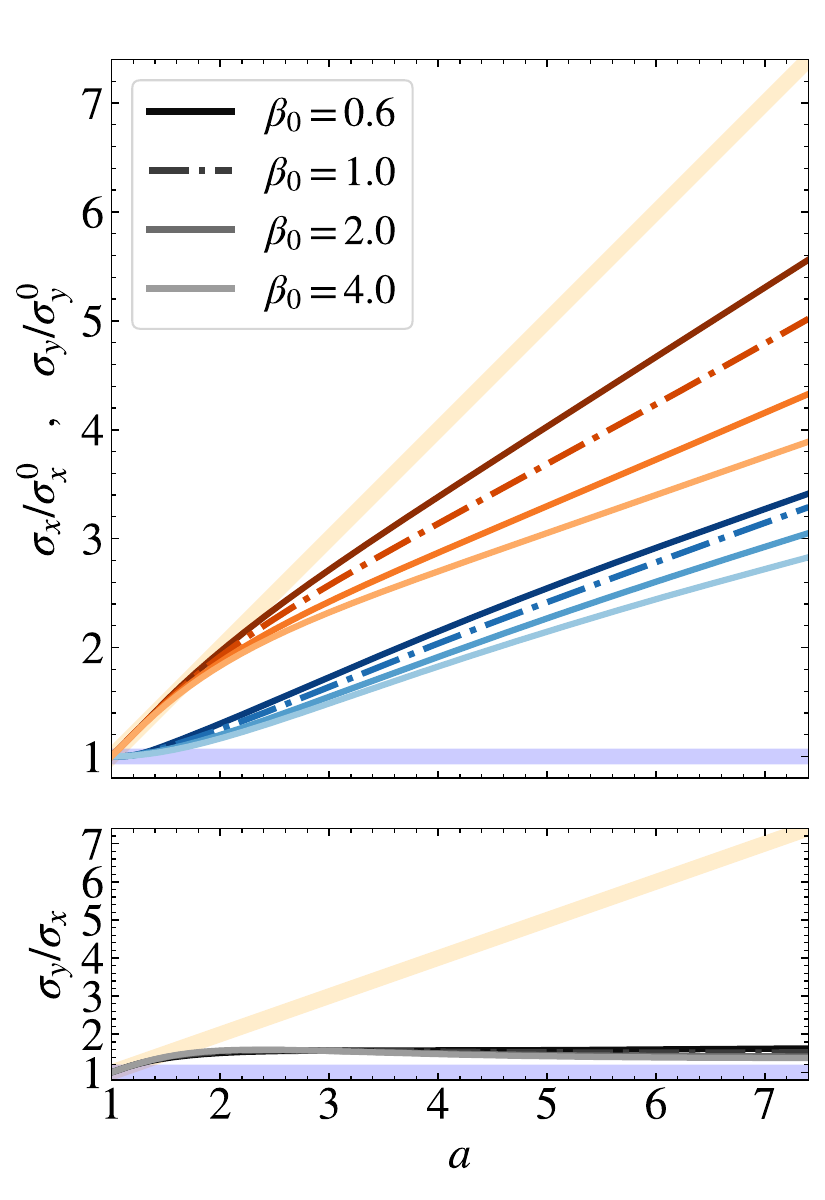}}
	\caption{
		Same as Fig.~\ref{fig-sigmas-eps0}, this time for runs B3, A3, C3, D3, that is, increasing $\beta_0$.
	}\label{fig-sigmas-beta}
\end{figure}
In Fig.~\ref{fig-sigmas-beta} we focus instead on the variation of the flux rope size associated with different plasma $\beta_0$ (the graphical conventions are the same as in Fig.~\ref{fig-sigmas-eps0}, this time lighter shades represent increasing $\beta_0$ values).
For increasing $\beta_0$, both the radial and transversal sizes decrease (top panel) and the aspect ratio is essentially invariant (bottom  panel); the ratio between magnetic tension and expansion timescales is kept constant, and a larger $\beta_0$ implies an isotropic increase of the gas (external) pressure gradients, thus the size of the flux rope increases less, independently of the directions.
However, radial expansion is clearly present even for run D3, which has $\beta_0 \sim 4$.
\section{Results: Turbulent simulations}
\label{section-results_turb}
We now consider a flux rope that is not isolated but embedded in turbulent fluctuations that are added in the whole domain (including the flux rope itself).
\subsection{Average and spectral quantities}
\begin{figure}[]
	\resizebox{\hsize}{!}{\includegraphics{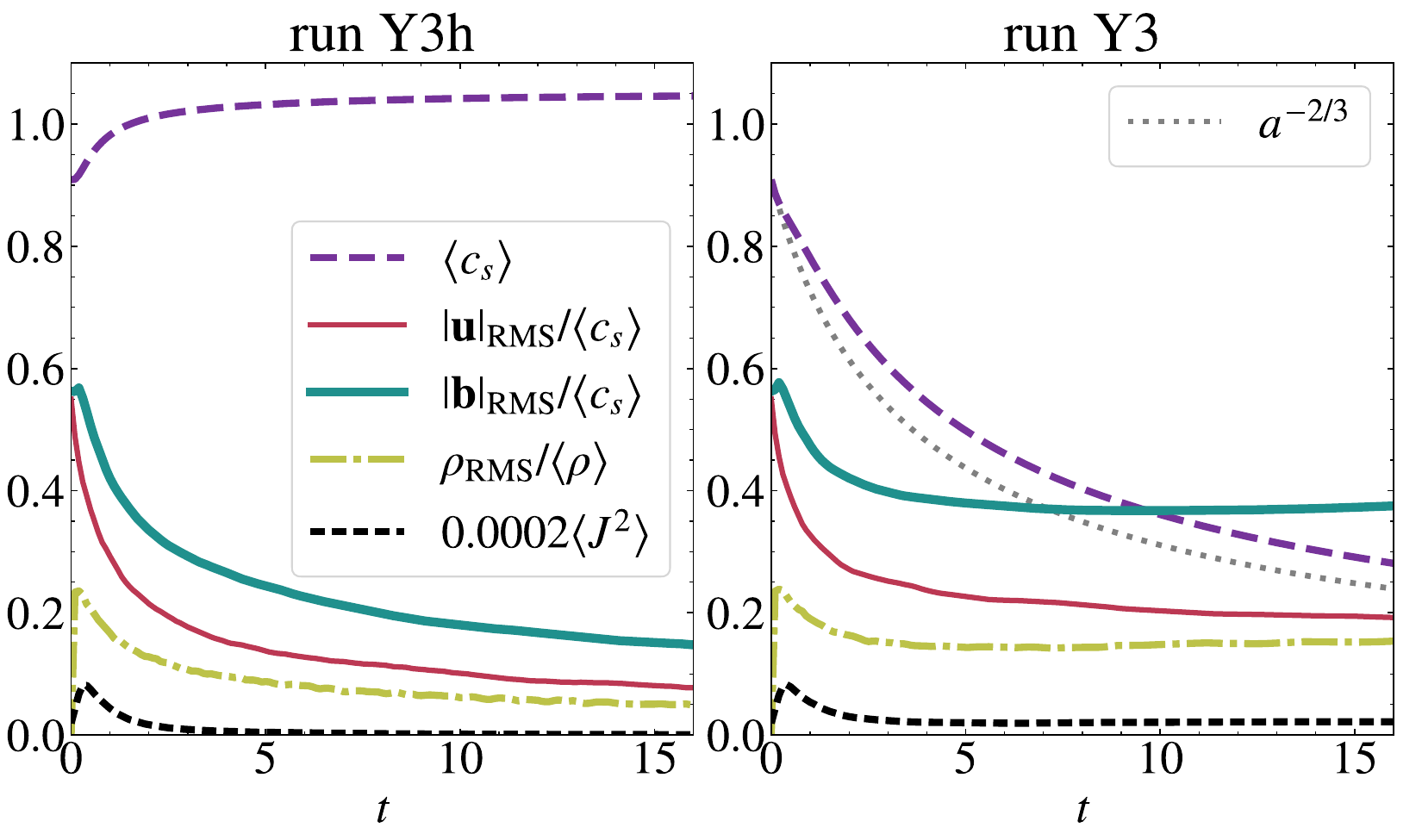}}
	\caption{
		Root mean square amplitude of fluctuations for runs Y3h (no expansion, left panel) and Y3 ($\eps_0=0.4$, right panel). Magnetic (green) and velocity (red) fluctuations are normalised to the average sound speed (magnetic fluctuations are expressed in Alfv\'en units before normalisation, see text), while density fluctuations (light green, dash-dotted) are normalised to the average density. The average sound speed and the square of the current density are also plotted with purple and black dashed lines, respectively. In the right panel, the predicted $a^{-2/3}$ decay for sound speed is drawn as a grey dotted line.
	}\label{fig-turb-rms}
\end{figure}

In Fig.~\ref{fig-turb-rms} we compare the temporal evolution of mean quantities for runs Y3h and Y3, the former having no expansion, the latter having the same $\eps_0 = 0.4$ as run A3.
In each panel we plot the average sound speed $\langle{c_\mathrm{s}}\rangle$, the velocity and magnetic RMS fluctuations, the relative density RMS fluctuations, and the square of the current density $\langle{|\mathbf{J}|^2}\rangle$.
Velocity and magnetic fluctuations are divided by the sound speed at each time, with magnetic fluctuations expressed in Alfv{\'e}n units, $\delta{b}=\delta{B}/\sqrt{\langle\rho\rangle}$.

We consider first the non expanding case (left panel). Despite the presence of the flux rope, fluctuations are almost at equipartition at the initial time. Since fluctuations are initially excited at relatively large wave-numbers, turbulence develops very quickly and intense small-scale currents form very soon (recall $\chi\gg1$), heating the plasma.
Hence the rapid decay of the normalised velocity and magnetic fluctuations is due to both the development of turbulence that drains energy from fluctuations and to the increase of the sound speed.
At later times $t\gtrsim5$, turbulence slows down, turbulent dissipation decreases, fluctuations decay more gently, and the sound speed only slightly increases.
Fluctuations are subsonic initially and remain largely subsonic at later time, with $\delta{u}/c_\mathrm{s}~\gtrsim\delta\rho/\langle\rho\rangle$.

When expansion is switched on (right panel), the evolution at early times remains basically unchanged. In other words, the development of turbulence in response to the out-of-equilibrium initial conditions is the same and more rapid than the effects of expansion (recall $\eps_\mathrm{T} \ll 1$).
Both the average density and temperature decrease according to Eqs.~\eqref{eqn-EBM-scal-n} and~\eqref{eqn-EBM-scal-T}, and the plasma heating now shows up as an initial less-than-adiabatic decrease of the sound speed.
After the initial phase, the sound speed follows an adiabatic decrease (grey dotted line), and the level of (normalised) velocity, magnetic, and density fluctuations settles to an approximately constant value that is larger than in the homogeneous case.
Thus, velocity and magnetic fluctuations (the latter in Alfv{\'e}n units) decay approximately as $a^{-2/3}$, while density fluctuations as $a^{-2}$. A small magnetic excess continues to develop at later times, possibly because of the persistence of the flux rope magnetic field.
\begin{figure}[]
	\centering
	\resizebox{\hsize}{!}{\includegraphics{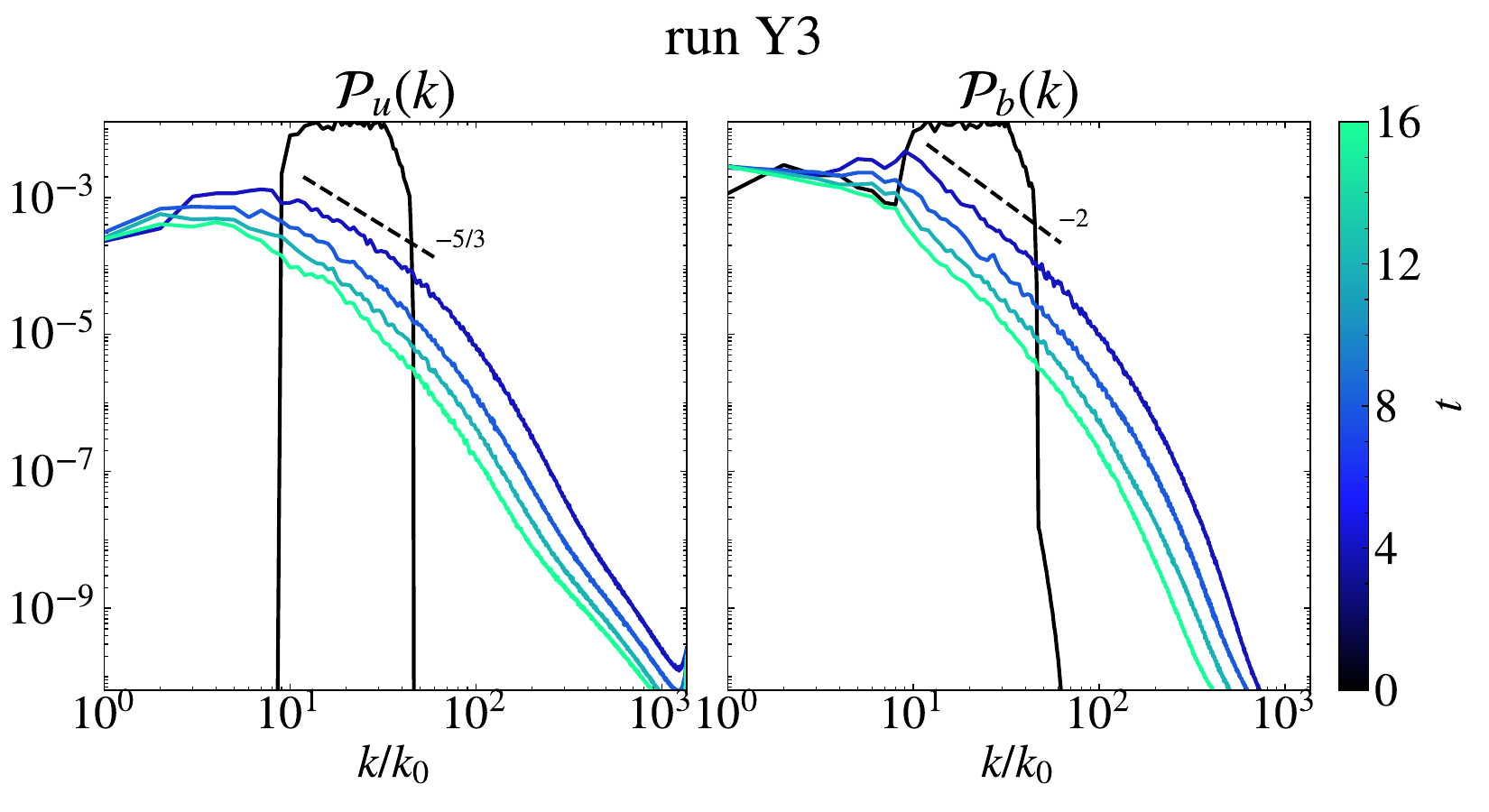}}
	\caption{
		Omnidirectional spectra of velocity and magnetic field (Eq~\eqref{eq-spectra-omni}) as a function of the normalised wave-number for run {Y3}. Different times are shown as indicated in the colour bar. The spectrum corresponding to the flux rope is visible at $k/k_0\lesssim6$ in the magnetic spectrum.
	}\label{fig-turb-spectra}
\end{figure}
\begin{figure*}[]
	\centering
	\resizebox{\hsize}{!}{\includegraphics{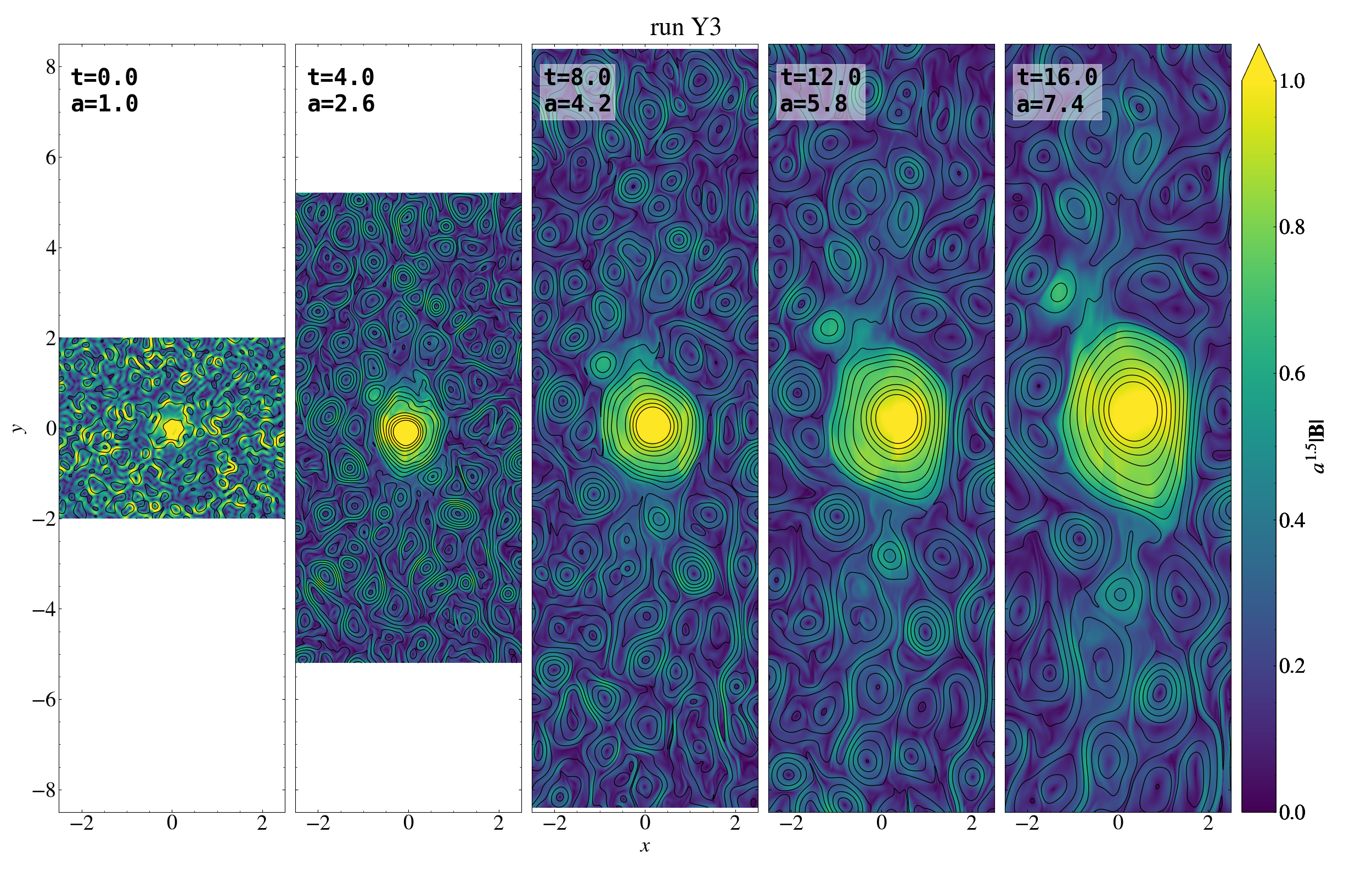}}
	\caption{
		Snapshots for successive times of the magnetic field for run {Y3}. The total magnetic field $|\mathbf{B}|$ is represented by the colour-coded map, and the in-plane magnetic field is shown with constant-$A_z$ black lines. Only a portion of the domain is shown, see text.
	}\label{fig-b-turb}
\end{figure*}
\begin{figure}[]
	\centering
	\resizebox{\hsize}{!}{\includegraphics{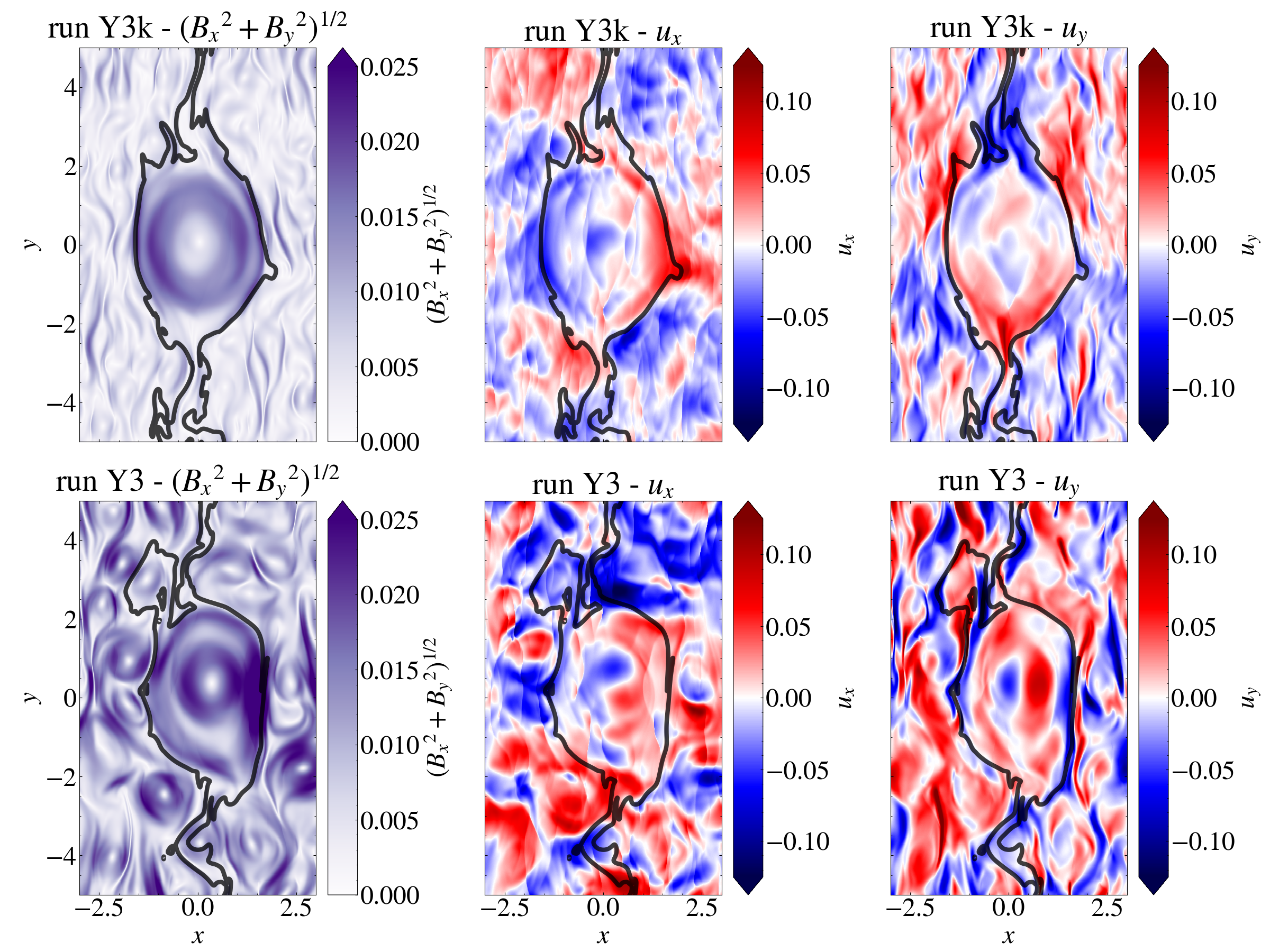}}
	\caption{
		Maps of the in-plane magnetic field, $B_\mathrm{pl} = \sqrt{{B_x}^2+{B_y}^2}$ (left panels, colour coded) and of the two components of the velocity field, $u_x$  and $u_y$ (central and right panels, colour coded), at $t=16$. The pinch tracer edge is drawn as a solid black line. Top and bottom represent respectively runs Y3k and Y3, differing in the excited scales of initial random fluctuations: run Y3k has $k\ge 4 k_\mathrm{FR}$ and run Y3 has $k\ge k_\mathrm{FR}$.
	}\label{fig-vxvy-turb}
\end{figure}
\begin{figure}[]
	\centering
	\resizebox{\hsize}{!}{\includegraphics{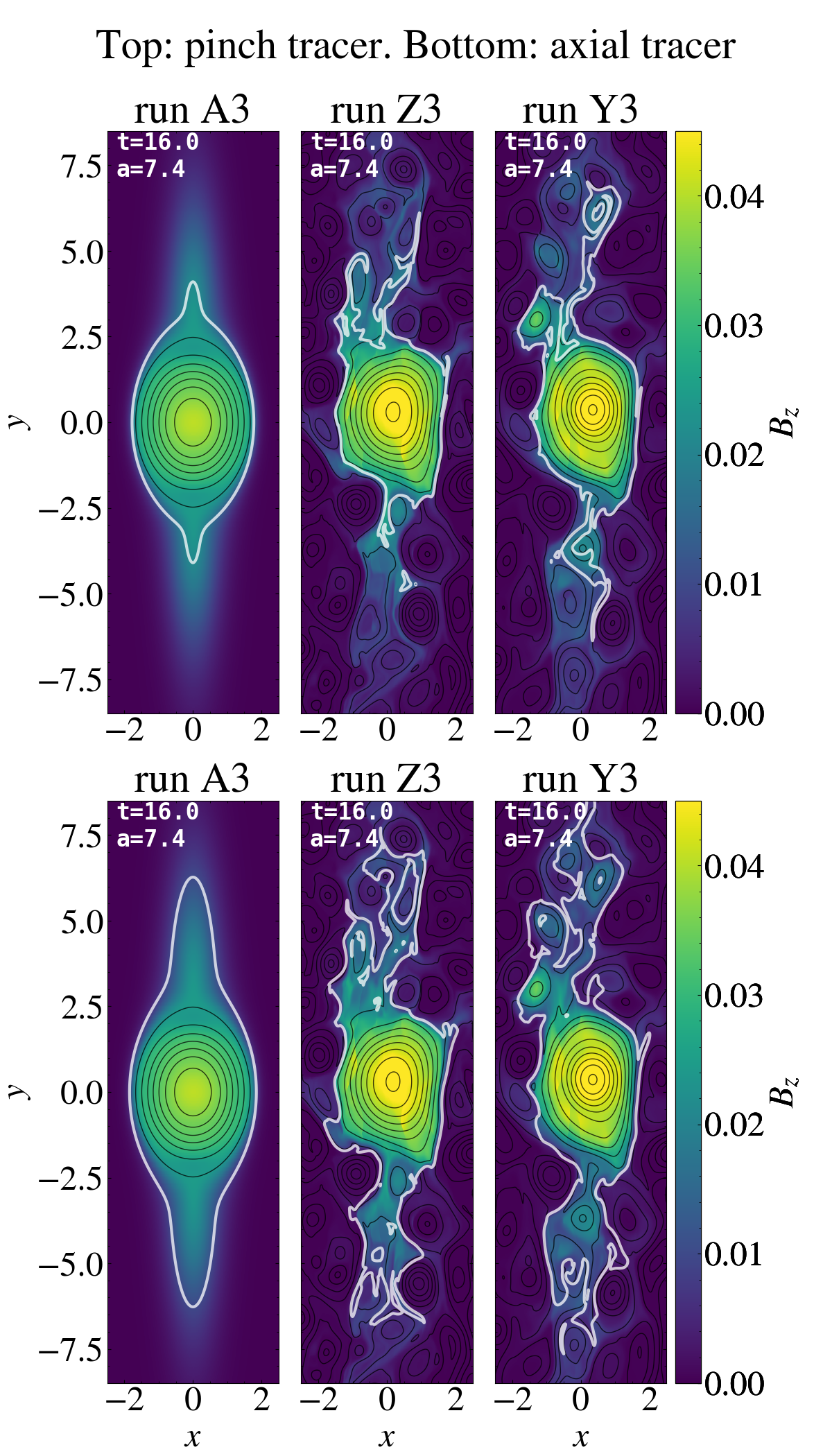}}
	\caption{
		Snapshot at $t=16.0$ ($a=7.4$) of the magnetic field for runs A3 (isolated flux rope), Z3 (with fluctuations smaller inside the flux rope by a factor 5), Y3 (with equal fluctuations amplitude everywhere). The axial field $B_z$ is represented by the colour-coded map, whereas the in-plane magnetic field is represented with constant-$A_z$ black lines. The two rows show the same fields (colour map and black iso-contours) but differ for the tracer that is plotted with a thick white line: in the top row we show the pinch tracer that initially is co-spatial with the in-plane (poloidal) magnetic field; in the bottom row the white lines bound the axial tracer that is initialised in the same region occupied by the out-of-plane (axial) magnetic field.
	}\label{fig-b-id-turb}
\end{figure}
\begin{figure}[]
	\centering
	\resizebox{\hsize}{!}{\includegraphics{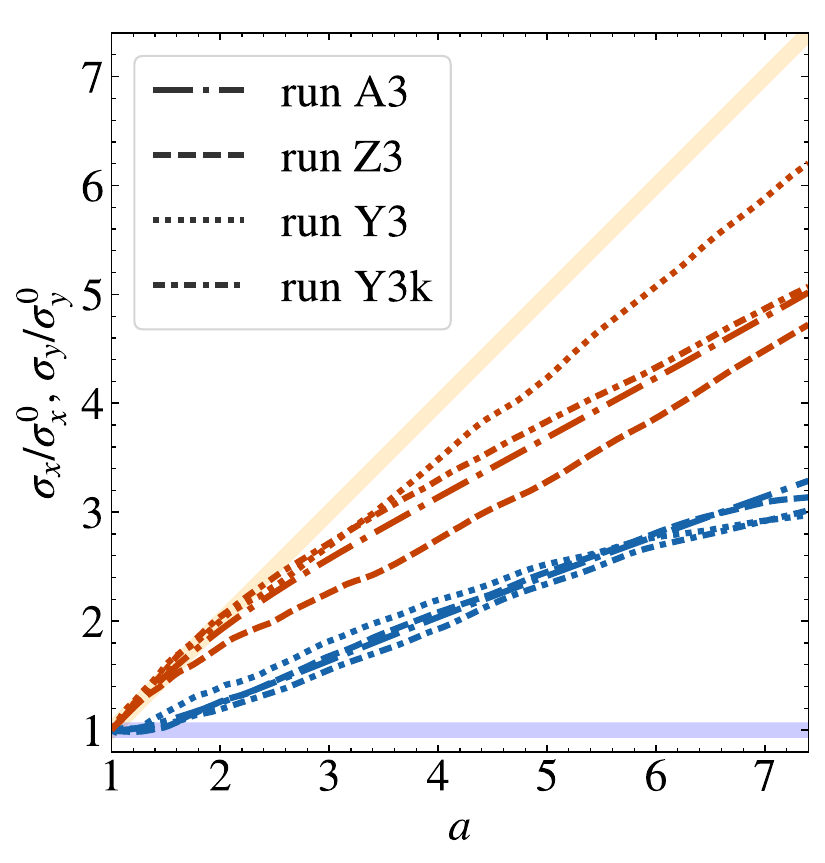}}
	\caption{
		Heliospheric evolution of the pinch tracer sizes ($\sigma_x$ and $\sigma_y$) for runs A3 (no turbulence, dash-dotted line), Z3 (turbulence out greater than in, dashed line), Y3 (turbulence in and out, dotted line), Y3k (turbulence in and out at smaller scales, densely dash-dotted line). The light blue (orange) broad line represents the expected ``kinematic'' trend along $x$ ($y$).
	}\label{fig-sigmas-turb}
\end{figure}

For the expanding simulation Y3, the evolution of the omnidirectional spectra of velocity and magnetic fluctuations are shown in Fig~\ref{fig-turb-spectra}, the latter again expressed in Alfv{\'e}n units, with lighter colours indicating successive times: $t = 0.0, 4.0, 8.0, 12.0, 16.0$, corresponding to an increase of the distance and transverse size up to a factor $7.4$.
In the plot, the abscissa is further normalised to the smallest wave-number, $k_0 = 2\pi/L_x$, which is the same at all times.
Initially (black line), the spectrum of velocity is flat and made of turbulent fluctuations only, while the magnetic field spectrum clearly shows the flux rope energy at $k\le6$.
In both panels the secular decay due to the EBM frictional forces is apparent, with spectra decreasing as a whole at successive times, and faster for $\vec{u}$ than for $\vec{b}$ (cf. Eqs.~\eqref{eqn-EBM-scal-u}-\eqref{eqn-EBM-scal-B}).
Velocity fluctuations of small amplitude develop at large scales ($k/k_0\lesssim6$), while in the magnetic field spectrum this range is dominated by the flux rope that remains as a distinguishable feature at all times.
At smaller scales, $15\lesssim{k/k_0}\lesssim70$, an inertial range develops with a power-law range, with a spectral slope slightly steeper than $-5/3$ for $\vec{u}$ and slightly steeper than $-2$ for $\vec{b}$.
In the magnetic spectrum the boundary between the flux rope and turbulent fluctuations at $k/k_0\approx6$ is smoothed out for large distances; this suggests a cross-scale interaction between turbulent vortexes and the flux rope.
Finally, for larger wave-numbers the spectra drop significantly because of the stretching of the domain:
at $t=4$ the largest transverse wave-vector is $k_y^{max}/k_0 = 1024/2.6~\approx 394$, and in the last spectrum it further reduces to $k_y^{max}/k_0\approx138$.
In other words, as the distance increases the omnidirectional spectra at the smallest scales are made mainly of radial wave-vectors.
\subsection{Radial expansion and ageing of turbulence}
We now focus on the evolution of run Y3 in real space:
the heliocentric evolution of the magnetic field intensity and field lines for run Y3 is shown in Fig.~\ref{fig-b-turb} with the same style and conventions as Fig.~\ref{fig-maps}.
The snapshots are taken at the same times of the omnidirectional spectra and span the whole duration of the simulation (unlike Fig.~\ref{fig-maps} which only covered $t \leq 8.0$), that is, $t = 0.0, 4.0, 8.0, 12.0, 16.0$ and a final domain stretching of $7.4$.
For better visual representation, $|\mathbf{B}|$ is compensated by $a^{3/2}$ and the 2D maps only show a portion of the domain, $x \in (-2.5, +2.5)$ and $y \in (-8.5,+8.5)$, with the actual domain boundaries being located at $x = \pm 4$ at all times, and at $y=\pm14.8$ at the final time.
The random fluctuations quickly interact to form currents and turbulent vortexes that are about one third of the flux rope size in this initial phase.
Turbulent vortexes are expanding as well and lose energy because of the turbulent cascade, so that on average their size decreases with respect to the flux rope.
Around the boundaries of the flux rope (especially the radial ones) a number of small scale structures form that have isotropic shapes (e.g. vortex-like structures on the left) or vertical elongation (e.g. sheet-like structures on the right).
Comparison of the total magnetic field (colour map) and in-plane magnetic field (black iso-contour) also reveals that the out-of-plane $B_z$ is more intense in vortexes that are close to the flux rope's vertical edge.
Since 2D turbulence is unlikely to develop a strong out-of-plane magnetic field, this must originate from the diffusion and/or transport of the axial field of the flux rope (for example in the small blob on the top left corner of the flux rope). We will come back to this point later in this section.

The most striking fact emerging from Fig.~\ref{fig-b-turb} is that despite the strong turbulent field, the core of the flux rope remains clearly visible as the most intense magnetic field; as in the isolated case, the flux rope radial size grows, and its transverse size grows less than geometrically.
We expect departures from the isolated case to be controlled by the age of expanding turbulence, or its inverse $\eps_\mathrm{T}$.
In Fig.~\ref{fig-vxvy-turb} we compare the final state ($t = 16.0$) of two runs with different $\eps_\mathrm{T}$: run Y3k with $\eps_\mathrm{T} = 0.03$ in the top row, and run Y3 with $\eps_\mathrm{T} = 0.12$ in the bottom row. The in-plane magnetic field intensity $B_\mathrm{pl} = {({B_x}^2+{B_y}^2)}^{1/2}$ is shown in the left column, whereas the radial and transversal velocities $u_x$ and $u_y$ are shown in the central and right columns, respectively. In all panels, the pinch tracer is shown as a solid black line to indicate the flux rope boundaries; again, only a portion of the domain is shown for better visualisation.
Let us focus on run Y3k (top panels) first. The in-plane magnetic field shows turbulent fluctuations that are filamented and about $10$ times smaller in size and about $5$ times smaller in amplitude than the flux rope field.
The flux rope, as identified by the pinch tracer, is well bounded by the in-plane magnetic field in the radial direction. Instead, in the transverse direction its material is spread vertically, possibly because of diffusion outside the region bounded by magnetic tension.
In the top right and central panels, the velocity profiles causing radial expansion and transversal contraction can be recognised as in Fig.~\ref{fig-maps}.
In addition, the vertical flows cause velocity shears $u_y(x)$ at the radial edges of the flux rope that in turn lead to an enhancement of the in-plane magnetic field, clearly visible in the top left panel.
Summarising, at late times the kinematics is similar to the isolated case (run A3, Fig.~\ref{fig-maps} (b) and (c)) and turbulent fluctuations add a little perturbation to the flux rope structure. This is consistent with its $\eps_\mathrm{T} \ll 1$: the eddy turnover time is initially much shorter than the expansion time, so turbulence decays quickly and at later times few energy is left to perturb the flux rope, and the geometrical stretching forces the weakly interacting eddies into filamentary shapes.

The situation is different for our fiducial case, run Y3 (bottom panels of Fig.~\ref{fig-vxvy-turb}).
Turbulent fluctuations have a magnetic field intensity comparable to that of the flux rope, their size is relatively large (about $1/3$ of the flux rope), and have mainly vortex-like shapes.
A central core forms well inside the flux rope material identified by the tracer, and is surrounded by magnetic field enhancements at the flux rope edges.
As before, dispersion in the vertical direction takes the form of filamentary structures at the top and bottom edges, but also vortex-like structures are present (e.g. in the top left corner).
In the middle and right panels, the turbulent $\delta{u_{x,y}}$ are comparable in size and magnitude to those of the flux rope, with clear vortex-like patterns.
The ratio $\eps_\mathrm{T} \lesssim 1$ indicates that turbulence is dynamically relevant also at late times.
The central core of the flux rope rotates counter-clock wise, while now velocity shears appear at both the radial and transverse edges.
While it is clear that the transverse component of velocity has lost its dipolar structure at the vertical edges (right panel of Fig.~\ref{fig-vxvy-turb}), with
the aid of the tracer in the middle panel one can recognise a profile of the radial velocity $u_x(x)$ resembling the unperturbed case.
Thus, large-scale and large-amplitude turbulence is effective in perturbing the flux rope structure and dynamics along the transverse directions. Along the radial direction, the flux rope maintains approximately the radial velocity profile that was found in the isolated case and that defines radial expansion.
Clearly both $\chi$ and $\eps_0$ control the impact of turbulence on the flux rope equilibrium, suggesting that values of $\eps_\mathrm{T}$ closer to unity are more effective in perturbing the flux rope equilibrium; a better handle on the final state is obtained by controlling separately $k_\mathrm{turb}/k_\mathrm{FR}$ and $\delta B/B_{\theta,0}$.
\subsection{Magnetic field dispersion and size estimates}
We now focus on the causes of the dispersion of the flux rope material in the vertical direction, and in particular on the role of transport and magnetic diffusion.
On the one hand, transport is possible without diffusion because our initial equilibrium has a non-negligible axial field beyond the flux rope pinch (see Sect.~\ref{section_methods-init} and Fig.~\ref{fig-fluxrope-init}).
On the other hand, velocity maps display clear signatures of magnetic reconnection between the field of the flux rope and that of the vortexes (we note however that this process could be amplified by the presence of strong turbulent fluctuations inside the flux rope, arbitrarily set to $\delta{B}/B_{\theta,0}=1$ in this work).
To clarify the importance of both effects we compare in Fig.~\ref{fig-b-id-turb} the intensity map of the axial magnetic field at the final time for runs A3, Z3 and Y3, corresponding to an isolated case, a turbulent case with weaker fluctuation inside the flux rope, and uniform fluctuations.
For easier comparison with the last panel in Fig.~\ref{fig-b-turb}, on the colour map we superpose the iso-contour of the in-plane magnetic field with black lines; moreover, to identify the flux rope material, we draw the boundary of the pinch and the axial tracers in the top and bottom panels, respectively.
We recall that the pinch tracer is initially limited to the region bounded by the in-plane magnetic field, while the axial tracer extends further out in the vertical direction to cover also the out-of-plane flux rope magnetic field (see Figs.~\ref{fig-fluxrope-init} and \ref{fig-maps}).

We first focus on the isolated case (left column). The axial field $B_z$ extends well beyond the pinch tracer but is well bounded by the axial tracer.
This axial field is transported away from the flux rope in the vertical direction because the plasma that initially lies outside the in-plane flux rope field is freely stretched in that direction without the restoring force provided by the magnetic tension.
Thus, no magnetic diffusion is at work in the isolated case.
When turbulent fluctuations are present, but with a smaller amplitude inside the flux rope than outside of it (middle panels), the axial field has a similar vertical extent, a more irregular pattern due to the turbulent motions, and about the same intensity.
Now, the pinch tracer cover most of the axial field in the vicinity of the flux rope, indicating that this intense $B_z$ originates from diffusion (reconnection) and subsequent transport.
Finally, the right panels show that if turbulent fluctuations are more intense inside the flux rope, diffusion is more effective and intense $B_z$ can be transported further away, reaching distances comparable to the flux rope size.
In the radial direction the two tracers are basically co-spatial, indicating that reconnection and transport are effective only in the transverse direction.

Finally, we show the heliocentric evolution of radial and transverse sizes as measured with the pinch tracer for the turbulent runs, with run A3 as a reference.
In Fig.~\ref{fig-sigmas-turb} we plot again $\sigma_x$ and $\sigma_y$ as a function of distance for runs A3 (loosely dash-dotted), Y3k (densely dash-dotted), Y3 (dotted) and Z3 (dashed), with the same colour notation as in Fig.~\ref{fig-sigmas-eps0}.
As anticipated, the increase of the radial size is basically unchanged (all the lines are very close to each other): turbulence has a little effect despite its $\delta B \approx B_{\theta,0}$ and irrespective of the internal level of turbulence.
On the contrary, the transverse size of the flux rope is strongly affected when the turbulence is durable (runs Y3 and Z3, dotted and dashed lines), approaching the kinematic trend $\sigma_y \propto a$ in the more extreme case of run Y3; on the contrary, the short-lived turbulence of run Y3k shows little variation.
We note, however, that the measure of the flux rope size, $\sigma_x$ ($\sigma_y$), is obtained by averaging in the $y$ ($x$) direction, thus it smears out the differences that are instead visible in the 2D maps of Fig.~\ref{fig-b-id-turb}
and could be grasped by cutting the structure at its centre (which is more similar to what a spacecraft might see).
%
%
\section{Comparison with previous works and discussion}
\label{section-discussion}
\begin{figure}[]
	\centering
	\resizebox{\hsize}{!}{\includegraphics{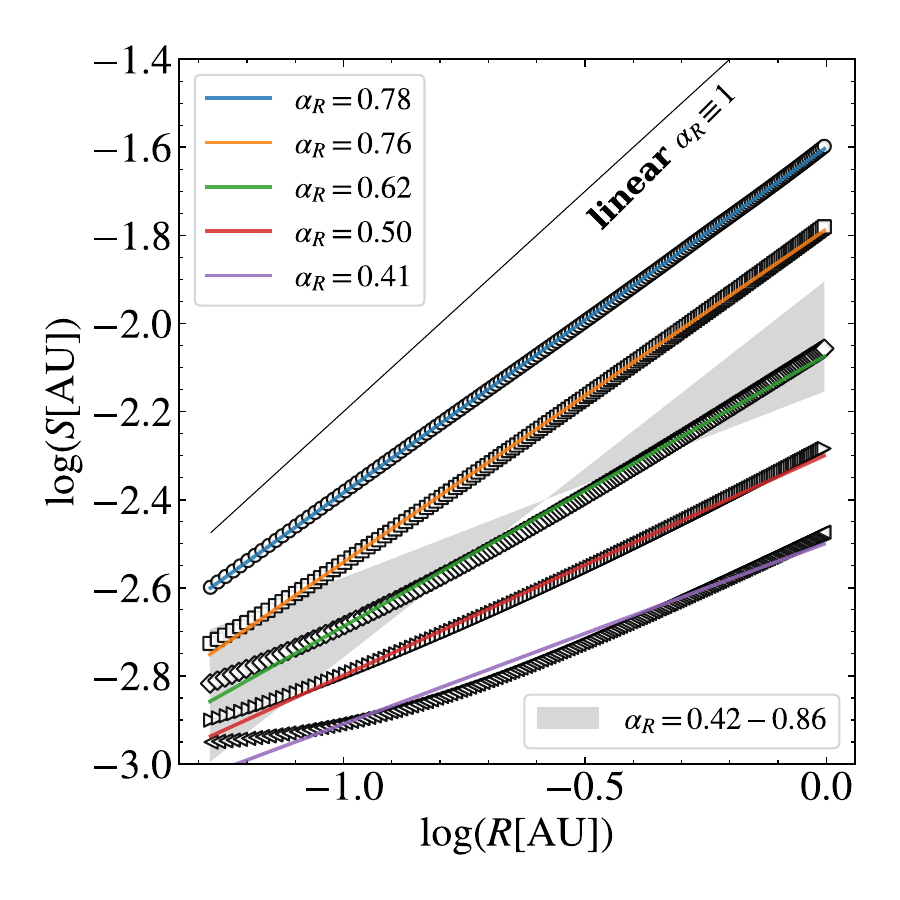}}
	\caption{
		Heliospheric evolution of the flux rope radial size $S[AU]$ versus distance $R[AU]$ for runs A2-A6. Only the range $a>2$ ($R>\SI{0.29}{AU}$) is considered, to avoid the first dynamical transient for most runs (even though it is still visible and important in runs A4-A6). The estimates of $S$ using $\sigma_x$ are drawn as markers (different shapes for different runs), and each run has been fitted with a power-law (solid lines with different colours). The solid thin black line is the linear ($\alpha_R\equiv1$) trend. The shaded grey area highlights the range between the minimum and maximum $\alpha_R$ values obtained by randomly sampling data points from the combined set of runs A2-A6 (see text for details).
	}\label{fig-S-R}
\end{figure}
\subsection{General features and comparison at 1AU}
\begin{table*}
	\caption{Comparison between averages based on superposed epoch analysis at $\SI{1}{AU}$ \citep[][Table 1, ``Cat-III'' set]{salman2020ApJ...904..177S} and our runs A3, AR and A6, corresponding to different values of the non-dimensional expansion rate $\eps_0$.}
	\label{table-1AU-comparison}
	\centering
	\begin{tabular}{l c c c c c c | c}
		\hline\hline
		{} & $\langle{B}\rangle[\SI{}{nT}]$ & $\langle{n}\rangle[\SI{}{cm^{-3}}]$ & $L_\mathrm{FR}[\SI{}{AU}]$ & $U_0 [\SI{}{km s^{-1}}]$ & $V_\mathrm{exp}[\SI{}{km s^{-1}}]$ & $\langle\beta\rangle$ & $\eps_0$ \\
		\hline
		Reference (1AU) & $8.0\pm3.1$ & $4.6\pm3.3$ & $0.240\pm0.125$ & $402\pm65$ & $17\pm30$ & $0.1\pm0.1$ & {} \\
		\hline
        run A3 (1AU)    & 4.0 & 1.9 & 0.16 & 180 & 13.7 & 0.30 & 0.4 \\
        run AR (1AU)   & 4.5 & 2.2 & 0.12 & 466 & 19.4 & 0.30 & 1.0 \\
        run A6 (1AU)    & 5.8 & 2.7 & 0.08 & 1440 & 35.7 & 0.25 & 3.2 \\
		\hline
	\end{tabular}
	\tablefoot{
		For all the runs we fixed the values of $B^0 = B_\mathrm{FR} = \SI{120}{nT}$, $\rho^0 = \rho_\mathrm{bg} = \SI{236}{} m_\mathrm{p} \SI{}{cm^{-3}}$ and $L^0 = L_\mathrm{FR} =  \SI{0.05}{AU}$. See the text for details.
	}
\end{table*}
Our simulations naturally reproduce a configuration with a flux rope at a lower temperature than the surrounding solar wind of the same velocity \citep[one of the main observational signatures of magnetic clouds, see e.g.][]{richardson1995JGR...10023397R}, and a decreasing plasma $\beta$ inside the flux rope with heliocentric distance. The rarefaction/cooling wavefront produced by the dynamical imbalance inside the flux rope accounts for the former, whereas the latter can be explained as a superposition of the spherical decay (magnetic pressure decreasing less than kinetic pressure) and the local advection motions (radial expansion) which stretch the magnetic field profile.

Previous studies \citep[][]{demoulin2008SoPh..250..347D, demoulin2009A&A...498..551D, gulisano2010A&A...509A..39G} usually found the different heliocentric scaling laws of magnetic and kinetic pressures to be the main culprit for the radial expansion; while this remains true for our analysis, our results suggest that the anisotropy and spatial dilation due to the spherical nature of solar wind flow also play a role, implying different behaviours for radial and transversal directions (affecting the section's final aspect ratio) and depending on the ratio between internal and external timescales (controlled here by the non-dimensional expansion rate $\eps_0$).

To relate our non-dimensional expansion rate $\eps_0$ to actual flux ropes and to compare our results to observed quantities at $\SI{1}{AU}$, we choose as a reference the superposed epoch analysis at $\SI{1}{AU}$ reported in \citet[][]{salman2020ApJ...904..177S}, Table 1, ``Cat-III'' set (CMEs with no sheath, closer to our assumption of constant propagation speed).
To obtain dimensional values from our runs,
we fix the dimensional units at $R_0 = 30 R_\sun$: the flux rope axial field $B^0 = B_\mathrm{FR} = \SI{120}{nT}$, the (uniform) mass density $\rho^0 = \rho_\mathrm{bg} = \SI{236}{} m_\mathrm{p} \SI{}{cm^{-3}}$ and the initial flux rope width $L^0 = L_\mathrm{FR} = \SI{0.05}{AU}$.
This leaves $\eps_0$ to be determined from the (constant) propagation speed $U_0$, or vice versa. For runs A3 and A6 we fix $U_0$ using the respective values of $\eps_0$ indicated in Table~\ref{table-runs};
then we also consider one more simulation, run AR, for which we choose $U_0 \simeq \SI{466}{km/s}$ \citep[average of the leading edge speeds distribution traced up to $\SI{32}{R_\sun}$, from][]{gopalswamy2010ASSP...19..289G} which implies $\eps_0 = 1.0$.
If we considered all the dimensional parameters to be fixed except from $U_0$, then the range of values of $\eps_0$ used in this study would correspond to constant radial speeds $U_0\sim\SI{90}{km/s}$ to $\SI{1500}{km/s}$.
It should be stressed once more that these are just examples: the same value of $\eps_0$ can describe very different magnetic clouds, but at the same time two extremely similar magnetic clouds with different ejection speeds will have very different values of $\eps_0$.

Table~\ref{table-1AU-comparison} shows in the first row our reference, and in the other rows the results from our runs.
We get values lower than average for magnetic field, density and radial size, and a higher than average plasma $\beta$; most of our estimates seem to fall within the standard deviations; a particularly good agreement is found for the expansion velocities, whereas the propagation velocity is clearly out of range because we assumed it constant.
The higher $\langle\beta\rangle$ and the smaller radial extent might be attributed to our limitations on the values of $\beta_0$ (see the next section for a further comment).
\subsection{Dimensionless estimates of radial expansion}
\begin{figure}[]
	\centering
	\resizebox{\hsize}{!}{\includegraphics{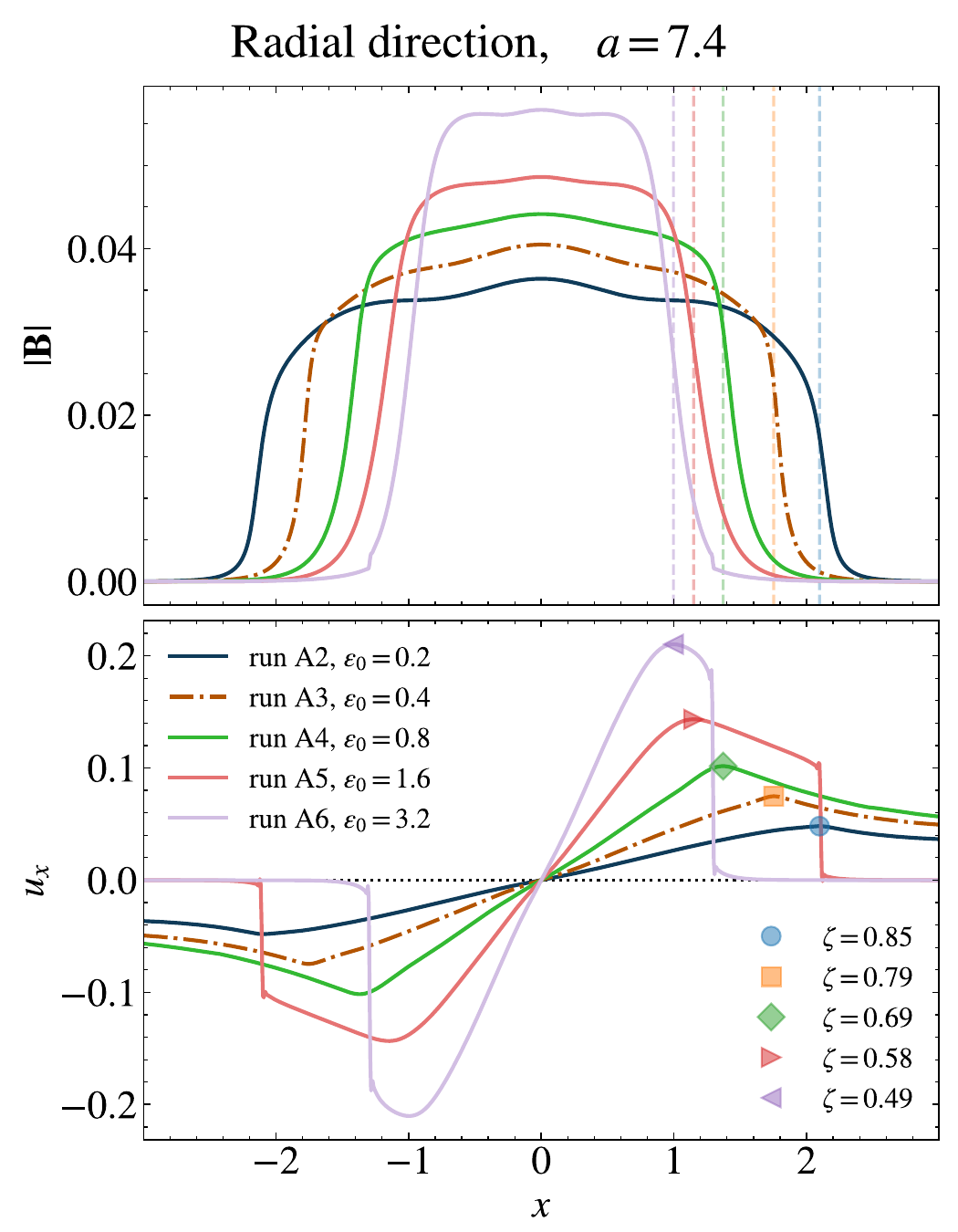}}
	\caption{
		Total magnetic field intensity $|\mathbf{B}|$ (top panel) and radial velocity $u_x$ (bottom panel) for runs A2-A6, along a radial cut through the flux rope axis, for the final times ($R\simeq\SI{1}{AU}$).
		In the bottom panel, the peak of $u_x$ is drawn with markers (the same as in Fig.~\ref{fig-S-R} for each run) and labelled with the corresponding estimate of the non-dimensional expansion parameter $\zeta$; the position corresponding to the peak is drawn as a light dashed vertical line in the top panels, to compare it to the magnetic boundaries.
		Run {A3} is shown as a dash-dotted line in both panels.
	}\label{fig-expparam1AU}
\end{figure}
Given the large variability of flux rope parameters, we can instead focus on dimensionless estimates.
From our trends of $\sigma_x(a)$ we can directly estimate the heliocentric scaling exponent of the flux rope's radial size, defined such that $S \propto R^{\alpha_R}$.
The $\sigma_x(a)$ we observe is made of an initial transient phase, coinciding with the span of non-zero radial dynamical imbalance (cf. the bottom left panel of Fig.~\ref{fig-acceleration-scalings}), whose duration in heliocentric coordinate depends on $\eps_0$ (Fig.~\ref{fig-sigmas-eps0}), and of a second asymptotic linear phase.
Because of the dynamical transient with little to no radial expansion, the effective scaling exponent is necessarily lower than its asymptotic value.
To consider only the latter phase of constant radial expansion, we fix the fitting interval as $a>2$ ($R>\SI{0.29}{AU}$); this works very well for runs A2-A3, but we still get a transitional phase for runs A4-A6, which leads to a systematic underestimation of $\alpha_R$.
Using this range, we find 
$\alpha_R = 0.78$ to $0.41$ for runs A2 to A6 ($\eps_0 = 0.2$ to $3.2$), indicating a clear anti-correlation between $\alpha_R$ and $\eps_0$.
As was expected from visual inspection of Fig.~\ref{fig-sigmas-eps0}, $\alpha_R < 1$ for all the considered cases.
We also draw as a grey shaded area the envelope between minimum and maximum slope obtained by randomly collecting $10$ samples from each run and using the ensemble to perform a ``statistical'' power-law fit, repeating for several random pickups. We can see that, based on which points we pick (i.e. on where we sample the different ``events'') the asymptotic slopes of single kinds of CMEs can be under or overestimated, possibly overestimating our highest $\alpha_R$ \citep[consistently with previous discussions, e.g. in][]{salman2024ApJ...966..118S}.

Most of our values of $\alpha_R$ are consistent with previous works, such as
$0.78\pm0.10$ \citep{bothmer1998AnGeo..16....1B},
$0.92\pm0.07$ \citep{liu2005P&SS...53....3L},
$0.61\pm0.09$ \citep{leitner2007JGRA..112.6113L},
$0.78\pm0.12$ \citep{gulisano2010A&A...509A..39G},
even though they fall on the lower end.

We also extract from our radial velocity profiles at $\SI{1}{AU}$ the non-dimensional expansion parameter $\zeta$, which gives a local estimate of the radial expansion exponent, and is defined following \citet{gulisano2010A&A...509A..39G}:
\[
\zeta
=
\frac{\Delta V_x}{\Delta t} \frac{D}{V_c^2}
\quad ,
\]
which, in our framework, is computed as
\begin{equation}\label{eqn-def-zeta}
\zeta
=
\frac{\Delta {u}_x}{\Delta {x}} \frac{a}{{\eps}_0}
\quad ,
\end{equation}
where $\Delta x = x_\mathrm{front} - x_\mathrm{back}$ and $\Delta{u}_x$ is the corresponding radial velocity difference.

The values of $\zeta$ estimated from runs A2-A6 at $\SI{1}{AU}$ are shown in Fig.~\ref{fig-expparam1AU} together with the 1D radial profiles of local radial velocity (bottom panel) and total magnetic field (top panel) obtained by cutting through the axis; this also shows once more that the radial velocity peak acts on the edge of the flux rope.
We find $\zeta = 0.85$ to $0.49$ for runs A2 to A6 ($\eps_0 = 0.2$ to $3.2$), anti-correlated to $\eps_0$ similarly to $\alpha_R$.
Moreover, for weaker expansion rates the final magnetic field is less intense and more spread, whereas for stronger expansion rates the magnetic field is more intense and compact; this can be easily seen comparing for example run A2 ($\eps_0 = 0.2$, blue lines) with run A6 ($\eps_0 = 3.2$, violet lines).
These values are positively correlated to the values of $\alpha_R$ discussed above; in particular, we find $\alpha_R \lesssim \zeta$ for all our runs, meaning that at least for our conditions the local measure of expansion based on the velocity slope ($\zeta$) seems to overestimate the actual heliocentric scaling exponent of the radial size of the flux rope ($\alpha_R$).
We note that an intrinsic anti-correlation exists between $\zeta$ and $\eps_0$ by definition in Eq.~\eqref{eqn-def-zeta}, but the same reasoning as Sect.~\ref{section-results_ideal_parameters} holds: the radial expansion velocity scales more weakly with $\eps_0$ than does the available time.

The range of $\zeta$ values we obtain for different EBM expansion rates is roughly consistent with the observational estimates of
$0.81\pm0.19$ by \citet{demoulin2010AIPC.1216..329D} and $0.91\pm0.23$ by \citet{gulisano2010A&A...509A..39G}, although we fall again on the lower end $\zeta\lesssim0.8$.

Finally, the flux rope's peak magnetic field intensity follows a heliocentric power-law decay $|\mathbf{B}|_\mathrm{max}\propto R^{\alpha_B}$ with exponent $\alpha_B = -1.66$ to $-1.49$ for runs A2 to A6 ($\eps_0 = 0.2$ to $3.2$); the range is quite narrow and the values are again anti-correlated with $\eps_0$: higher expansion rates imply a slower magnetic field decay, as can be seen in the top panel of Fig.~\ref{fig-expparam1AU}.
The validity of $|\alpha_B|/2$ as a proxy for $\alpha_R$ appears to be better for lower values of $\eps_0$ (i.e. $t_\mathrm{A} \ll t_\mathrm{exp}$).
Our values of $\alpha_B$ for $|\mathbf{B}|_\mathrm{max}$ are consistent with previous works based on observations, such as
$-1.64\pm0.40$ \citep{leitner2007JGRA..112.6113L},
$-1.34\pm0.71$ \citep{good2019JGRA..124.4960G},
$-1.41\pm0.49$ \citep{vrsnak2019ApJ...877...77V},
$-1.91\pm0.25$ \citep{salman2020JGRA..12527084S},
$-1.81\pm0.84$ \citep{lugaz2020ApJ...899..119L},
$-1.67\pm0.88$ \citep{zhuang2023ApJ...952....7Z}.
%
%
\section{Conclusions}
\label{section-conclusions}
\subsection{Summary and main results}
In this study we have performed MHD numerical simulations using the expanding box model to study the heliospheric evolution of the 2.5D section of a magnetic flux rope propagating with the spherically expanding solar wind.

We first considered just the flux rope in the spherical solar wind flow, and varied two control parameters: the non-dimensional expansion rate $\eps_0$ (how fast the flux rope propagates, compared to its internal Alfv{\'e}n times) and the effective plasma beta $\beta_0$ (how energetic the flux rope is, compared to the environment).

The spherical flow with its intrinsic anisotropy perturbs the flux rope, which reacts and develops a local dynamics: its radial size grows (action of magnetic pressure), its transversal size resists to the spherical stretching (action of magnetic tension).
The system eventually reaches a series of successive equilibria which include the local flows.

Stronger expansion rates imply a more dominant magnetic pressure but also faster transit and stretching, resulting in a less effective internal dynamics, a more elongated aspect ratio, and an evolution close to spherical expansion.
Lower beta values imply isotropically larger sizes because of the more predominant magnetic pressure.
The internal dynamics is thus relevant especially for low $\beta$ and weak expansion rates $\eps_0 = t_\mathrm{A} / t_\mathrm{exp} < 1$.

In the relevant case of weak expansion, we also superposed turbulent fluctuations to the isolated configuration, introducing another timescale, the non-linear time $t_\mathrm{NL}$.
The transversal resistance to stretching is strongly affected, whereas the radial size growth is almost unaltered.
The turbulent shearing and distorting motions transport the plasma (especially where magnetic tension is weaker), produce secondary structures and shape the flux rope at small scales.
Turbulence is effective in perturbing the flux rope only when it is long-lived with respect to the expansion timescale; energetic but short lived fluctuations have little effect. We suggest turbulence to be relevant when $\eps_\mathrm{T} = t_\mathrm{NL} / t_\mathrm{exp} \lesssim 1$.

We validated our simulations comparing the results to a superposed epoch analysis at $\SI{1}{AU}$ \citep[][Table 1, ``Cat-III'' set]{salman2020ApJ...904..177S}, finding a quite good agreement. However, in our simulations we underestimate plasma parameters such as density, magnetic field intensity and radial size (see Tab.~\ref{table-1AU-comparison}), probably due to our parameter limitations.
We also estimated the non-dimensional expansion parameter $\zeta$, from the radial expansion velocity, and the radial size scaling exponent $\alpha_R$, from the radial size increase with distance $S \propto R^{\alpha_R}$.
We find such local and global radial expansion estimates to be comparable, and both clearly anti-correlated to our non-dimensional expansion rate $\eps_0$.
The peak magnetic field scaling exponent $\alpha_B$ seems to work well as a proxy for $\alpha_R$ (especially when $t_\mathrm{A} \ll t_\mathrm{exp}$).
Estimating $\alpha_R$ from a set of position-size data points randomly sampled from different simulations clearly masks the single events, and can even overestimate (underestimate) the strongest (weakest) radial expansion exponent, just because of chance.

Our results seem consistent with the widely accepted view explaining radial expansion of magnetic clouds through the imbalance between magnetic and kinetic pressures \citep[e.g.][]{demoulin2009A&A...498..551D}; yet, this work gives more details on the internal dynamical processes and shows that radial expansion survives in presence of turbulence.

Finally, we stress that the dynamical imbalance which produces radial expansion in our simulations would still arise even for different kinds of initial configurations: a purely magnetic force-free equilibrium would have different scaling laws for magnetic pressure and tension, and a pressure-confined axial flux tube (no magnetic tension) would have different scaling laws for magnetic and kinetic pressures; this kind of evolution is thus a direct consequence of solar wind's spherically expanding geometry.

We expect our results to be valid mainly for magnetic clouds without shock fronts, that is, propagating at velocities close to the ambient solar wind.
\subsection{Limitations and perspectives}
Our estimates of the radial size scaling exponent and the non-dimensional expansion parameter both fall in the lower end of the ranges coming from observations (see Sect.~\ref{section-discussion}); both also show a less than linear ($\propto R$) expansion; this might be due to our limited $\beta$ range related to our choice of initial condition (see Sect.~\ref{section_methods-init}).
Since the radial expansion effect increases in our simulations for decreasing $\beta_0$ (see Sect.~\ref{section-results_ideal_parameters}), we expect our results to approach the upper end of the observed values when having lower $\beta_0$ values.

Also, assuming the flux rope to still be unperturbed at the starting radial position $R_0 = 30 R_\sun$ is probably not very realistic: we might be overestimating the duration of the dynamically imbalanced phase compared to the transit time, which means underestimating the duration of the successive phase of constant radial expansion, overall obtaining a radial size smaller than average at $\SI{1}{AU}$.

Moreover, even though we fixed the helical twist parameter of our initial configuration by arbitrarily fixing $B_{\theta,0} / B_{z,0} = 1/2$, we expect flux ropes with different degrees of twist to have a different evolution (final radial size and aspect ratio) because of the different magnetic pressure-to-tension ratio, introducing an additional parameter. A 3D extension of this study is probably needed to more completely assess such a dependence.

The estimate of an EBM-like expansion parameter $\eps_0$ (or its version considering the helical twist $\eps_{0,\mathrm{eff}}$, see Sect.~\ref{section-results_ideal_parameters}) for multiple in-situ magnetic clouds might be assessed in a subsequent study and would provide a further test for the general validity of our results.

Finally, the interaction with the interplanetary magnetic field and the presence of a velocity difference with solar wind, both neglected here, are expected to (strongly) perturb the evolution described in this work: the former would probably erode the magnetic cloud as it expands, also adding magnetic pressure all around it; the latter would likely produce a shock front and a sheath region, also sparking additional dynamics and higher dynamical pressure gradients, which might compress the flux rope in the radial direction and produce the well-known ``pancaking'' effect.
\begin{acknowledgements}
We acknowledge partial financial support from the European Union - Next Generation EU - National Recovery and Resilience Plan (NRRP) - M4C2 Investment 1.4 - Research Programme CN00000013 ``National Centre for HPC, Big Data and Quantum Computing'' - CUP B83C22002830001 and by the European Union - Next Generation EU - National Recovery and Resilience Plan (NRRP)- M4C2 Investment 1.1- PRIN 2022 (D.D. 104 del 2/2/2022) - Project ``Modeling Interplanetary Coronal Mass Ejections'', MUR code 31. 2022M5TKR2, CUP B53D23004860006.
We acknowledge the CINECA award under the ISCRA initiative, for the availability of high performance computing resources and support from the Class C project TUCME, ``Turbulent evolution of Coronal Mass Ejections'', code HP10C29IHP (PI: M.~Sangalli).
M.~S. wishes to thank S.~Antiochos for useful comments on an early presentation of this work; the authors also wish to thank L.~Del Zanna for useful discussions. The authors wish to thank the referee for their comments which helped improve the manuscript.
\end{acknowledgements}
\bibliographystyle{aa}
\bibliography{paper}
%
\begin{appendix}
\section{Initial equilibrium condition}
\label{appendix-init}
The local radial profiles $f(r')$ of the plasma parameters used in the initial equilibrium configuration described in Eq.~\eqref{eqn-init}, which is the same for all the simulations shown here, are the following (the primes are dropped for readability):
\begin{align}\label{eqn-appendix-init}
	& \rho(r) \equiv \rho_\mathrm{bg}
\quad ,
	\\
	& B_z(r) = B_{z,0} / \cosh(r/\Delta_z)
\quad ,
	\\
	& B_\theta(r) = B_{\theta,0} (r/\Delta_\theta)^4 \exp(4(1-r/\Delta_\theta))
\quad ,
	\\
	& T(r) = P(r) / \rho(r)
\quad ,
\end{align}
where
\begin{align}\label{eqn-appendix-init-P}
	P(r) & =
	\rho_\mathrm{bg} T_\mathrm{bg}
	- \frac{1}{2} \left( {B_{\theta,0}}^2 \left(\frac{r}{\Delta_\theta}\right)^8 e^{(8(1-r/\Delta_\theta))} + {B_{z,0}}^2 \frac{1}{(\cosh(r/\Delta_z))^2} \right)
	\notag\\
	& + {B_{\theta,0}}^2 \frac{7!}{8^8} e^{8(1-r/\Delta_\theta)} \sum_{m=0}^{7} \frac{(8r/\Delta_\theta)^m}{m!}
\quad ,
\end{align}
and where the parameters were chosen to be $\Delta_z = \Delta_\theta = \Delta$.
\section{Equations for numerical integration}
\label{appendix-equations}
Our simulation code integrates the fully compressible, viscous and resistive EBM equations using rescaled variables following the appendix of \citet{rappazzo2005ApJ...633..474R}, since such an approach allows one to integrate the out-of-plane magnetic potential $A_z$ instead of $(B_x,B_y)$, which ensures zero-divergence since we are in 2.5D geometry. In particular, instead of using the EBM equations as described in \citet{grappin1993PhRvL..70.2190G}, we rescale $\rho \mapsto \hat{\rho} = a^2 \rho$ in order to have unit-average density, and $\vec{u} \mapsto \vec{\hat{u}}$ with $\hat{u}_\parallel = u_\parallel$ and $\hat{u}_\perp = u_\perp / a$ and $\vec{B} \mapsto \vec{\hat{B}}$ with $\hat{B}_\parallel = a^2 B_\parallel$ and $\hat{B}_\perp = a B_\perp$, in order to have an induction equation formally equivalent to the homogeneous MHD. This allows for the use of a scalar potential $\hat{\phi} \equiv \hat{A}_z$  for the evolution of the (rescaled) in-plane magnetic field. The temperature $T$ is not rescaled, so that $\hat{T} \equiv T$. The EBM equations written here differ from those of \citet{rappazzo2005ApJ...633..474R} in that the effects of compression and heating are included in the temperature equation.
The full EBM equations for the rescaled variables thus read:
\begin{subequations}
	\label{eqn-EBM_num}
	\begin{eqnarray}
		\partial_t{\hat{\rho}}
		& = &
		-{\hat{\nabla}}\cdot{\left( \hat{\rho}\vec{\hat{u}} \right)}
		\\
		\partial_t{\hat{\vec{u}}}
		& = &
		-\left( {\vec{\hat{u}}}\cdot{\hat{\nabla}}\right)\vec{\hat{u}}
		-\frac{1}{\hat{\rho}} \Pb\hat{\nabla}\left( \hat{P} +
		\frac{1}{2}{\vec{\hat{B}}}\cdot{\Pa\vec{\hat{B}}} \right)
		+\frac{1}{\hat{\rho}} a^{-2}
		{\vec{\hat{B}}}\cdot{\hat{\nabla}}\vec{\hat{B}}
		\notag\\
		& &- 2\frac{\dot{a}}{a} \Pp \vec{\hat{u}}
		+ \left(\partial_t{\hat{\vec{u}}}\right)_\mathrm{diss}
		\\
		\partial_t{\hat{B}_z}
		& = &
		\left[\hat{\nabla}\times{\left({\vec{\hat{u}}}\times{\vec{\hat{B}}}\right)}\right]\cdot\vvers_z
		+ \left(\partial_t{\hat{B}_z}\right)_\mathrm{diss}
		\\
		\partial_t{\hat{\phi}}
		& = &
		-\left( {\vec{\hat{u}}}\cdot{\hat{\nabla}} \right) \hat{\phi}
		+\left(\partial_t{\hat{\phi}}\right)_\mathrm{diss}
		\\
		\partial_t{\hat{T}}
		& = &
		-\left( {\vec{\hat{u}}}\cdot{\hat{\nabla}}\right) \hat{T}
		- (\gamma-1) \hat{T} \left({\hat{\nabla}}\cdot{\vec{\hat{u}}}\right)
		\notag\\
		& &{- 2(\gamma-1)\frac{\dot{a}}{a} \hat{T}}
		+ \left(\partial_t{\hat{T}}\right)_\mathrm{diss}
	\end{eqnarray}
\end{subequations}
where the dissipation terms are defined as:
\begin{subequations}
	\label{eqn-EBM_num_diss}
	\begin{eqnarray}
		\left(\partial_t{\hat{\vec{u}}}\right)_\mathrm{diss}
		& = &
		\frac{\mu}{\hat{\rho}} \left[ \hat{\nabla}^2 \vec{\hat{u}}
		+\frac{1}{3}\hat{\nabla}\left({\hat{\nabla}}\cdot{\vec{\hat{u}}}\right) \right]
		\\
		\left(\partial_t{\hat{\vec{B}}}\right)_\mathrm{diss}
		& = &
		\eta \hat{\nabla}^2 \vec{\hat{B}}
		\\
		\left(\partial_t{\hat{\phi}}\right)_\mathrm{diss}
		& = &
		\eta\hat{\nabla}^2\hat{\phi}
		\\
		\left(\partial_t{\hat{T}}\right)_\mathrm{diss}
		& = &
		-(\gamma-1)
		\left[
		\frac{1}{\hat{\rho}}\Pa \vec{\hat{B}}\cdot\partial_t{\vec{\hat{B}}}|_\mathrm{d}
		+
		\Pc\vec{\hat{u}}\cdot\partial_t{\vec{\hat{u}}}|_\mathrm{d}
		\right]
		\notag\\
		& & +
		\frac{\kappa}{\hat{\rho}}\hat{\nabla}^2\hat{T}
	\end{eqnarray}
\end{subequations}
where the non-dimensional $\mu$ and $\kappa$ are in units of $\rho^0 (L^0)^2 / t^0$, the non-dimensional $\eta$ is in units of $(L^0)^2 / t^0$ and where the following $3\times3$ tensors are defined:
\begin{subequations}
	\begin{eqnarray*}
		\Pa &=& a^{-2}(\mathcal{I}-\Pp)+\Pp = \mathrm{diag}(a^{-2},1,1) \quad ,\\
		\Pb &=& (\mathcal{I}-\Pp)+a^{-2}\Pp = \mathrm{diag}(1,a^{-2},a^{-2}) \quad ,\\
		\Pc &=& (\mathcal{I}-\Pp)+a^2\Pp    = \mathrm{diag}(1,a^2,a^2) \quad .
	\end{eqnarray*}
\end{subequations}
An artificial diffusion term is also added to the passive scalar evolution to limit numerical instabilities:
\begin{equation}
	\partial_t s
	=
	-\hat{\vec{u}}\cdot\hnabla s
	+\xi\hnabla^2 s
\quad ,
	\label{eqn-EBM-PS_num}
\end{equation}
with a non-dimensional artificial diffusivity $\xi$.
\\
The functional form of the $\vec{\hat{u}}$ and $\vec{\hat{B}}$ dissipative terms is the same as in standard (non expanding) MHD equations. However, such terms are computed using the numerical gradients and act on the rescaled fields. Thus, unlike the ideal terms, the dissipative ones do not come from a formal derivation starting from MHD dissipative terms \citep[cf. section 2.2 of][]{montagud-camps2018ApJ...853..153M}. Moreover, the dynamic viscosity $\mu$, magnetic resistivity $\eta$ and thermal diffusivity $\kappa$ can be imposed to decrease with time/heliocentric distance; by doing so we can achieve the highest possible effective resolution without over-damping the fluctuations. Energy conservation is however self-consistent, since the same terms that dissipate kinetic and magnetic energy (including a possible dependence $\mu(R){=}\eta(R){=}\kappa(R)$) are included as cooling/heating sources in the temperature equation, with anisotropic scaling factors (tensors $\Pa$ and $\Pc$) so as to keep the total physical energy density constant throughout the simulation.
\end{appendix}
\end{document}